\documentclass{article}

\usepackage{PRIMEarxiv}

\usepackage[utf8]{inputenc} 
\usepackage[T1]{fontenc}    
\usepackage{hyperref}       
\usepackage{url}            
\usepackage{booktabs}       
\usepackage{amsfonts}       
\usepackage{nicefrac}       
\usepackage{microtype}      
\usepackage{lipsum}
\usepackage{fancyhdr}       
\usepackage{graphicx}       
\graphicspath{{figs/}}     

\usepackage[table,xcdraw]{xcolor}
\definecolor{newcolor}{rgb}{.8,.349,.1}

\usepackage[switch,pagewise]{lineno} 

\usepackage{subcaption}
\usepackage{xspace}

\usepackage{hyperref}
\usepackage{multirow}
\usepackage[justification=centering]{caption}
\usepackage{booktabs}
\usepackage{comment}
\usepackage[utf8]{inputenc}

\newcommand{\ie}{\emph{i.e.,}\xspace}
\newcommand{\eg}{\emph{e.g.,}\xspace}

\newcommand{\cv}{COVID-19\xspace}
\newcommand{\PUBDATASET}{COVID-CT-MD\xspace}

\graphicspath{{figs/}{../}}

\title{COVID-VR: A Deep Learning COVID-19 Classification Model Using Volume-Rendered Computer Tomography
}

\author{
  Noemi Maritza L. Romero, Ricco V. Soares, Mariana R. Mendoza, João L. D. Comba\\
  Instituto de Informática \\
Universidade Federal do Rio Grande do Sul - UFRGS \\
  Porto Alegre\\
  \texttt{\{nmlromero, ricco.soares, mrmendoza, comba\}@inf.ufrgs.br} \\
}

\begin{document}
\maketitle

\begin{abstract}
The COVID-19 pandemic presented numerous challenges to healthcare systems worldwide. Given that lung infections are prevalent among COVID-19 patients, chest Computer Tomography (CT) scans have frequently been utilized as an alternative method for identifying COVID-19 conditions and various other types of pulmonary diseases. Deep learning architectures have emerged to automate the identification of pulmonary disease types by leveraging CT scan slices as inputs for classification models.
This paper introduces COVID-VR, a novel approach for classifying pulmonary diseases based on volume rendering images of the lungs captured from multiple angles, thereby providing a comprehensive view of the entire lung in each image. To assess the effectiveness of our proposal, we compared it against competing strategies utilizing both private data obtained from partner hospitals and a publicly available dataset. The results demonstrate that our approach effectively identifies pulmonary lesions and performs competitively when compared to slice-based methods.
\end{abstract}

\keywords{COVID-19 \and Deep Learning \and Classification Models \and Computer Tomography \and Volume Rendering}

\section{Introduction}

The \cv pandemic began spreading extensively in early 2020, resulting in the unfortunate loss of millions of lives worldwide. Accurate and timely diagnosis of \cv is crucial, especially during periods of high demand that can strain healthcare systems. The gold standard method for diagnosing \cv is the reverse transcriptase-polymerase chain reaction test (RT-PCR)~\cite{Hernandez-Huerta2020-yx}. However, RT-PCR can be time-consuming and yield false-negative results ~\cite{Pecoraro2021-gn,HASSAN2022105123}. An alternative approach involves analyzing chest images obtained from X-rays or Computer Tomography (CT) scans~\cite{Tabik2020,Roberts2021}. 
Due to their higher resolution, CTs can lead to more precise diagnostics. 

\begin{figure*}[!tbh]
    \centering
    \includegraphics[width=0.95\linewidth]{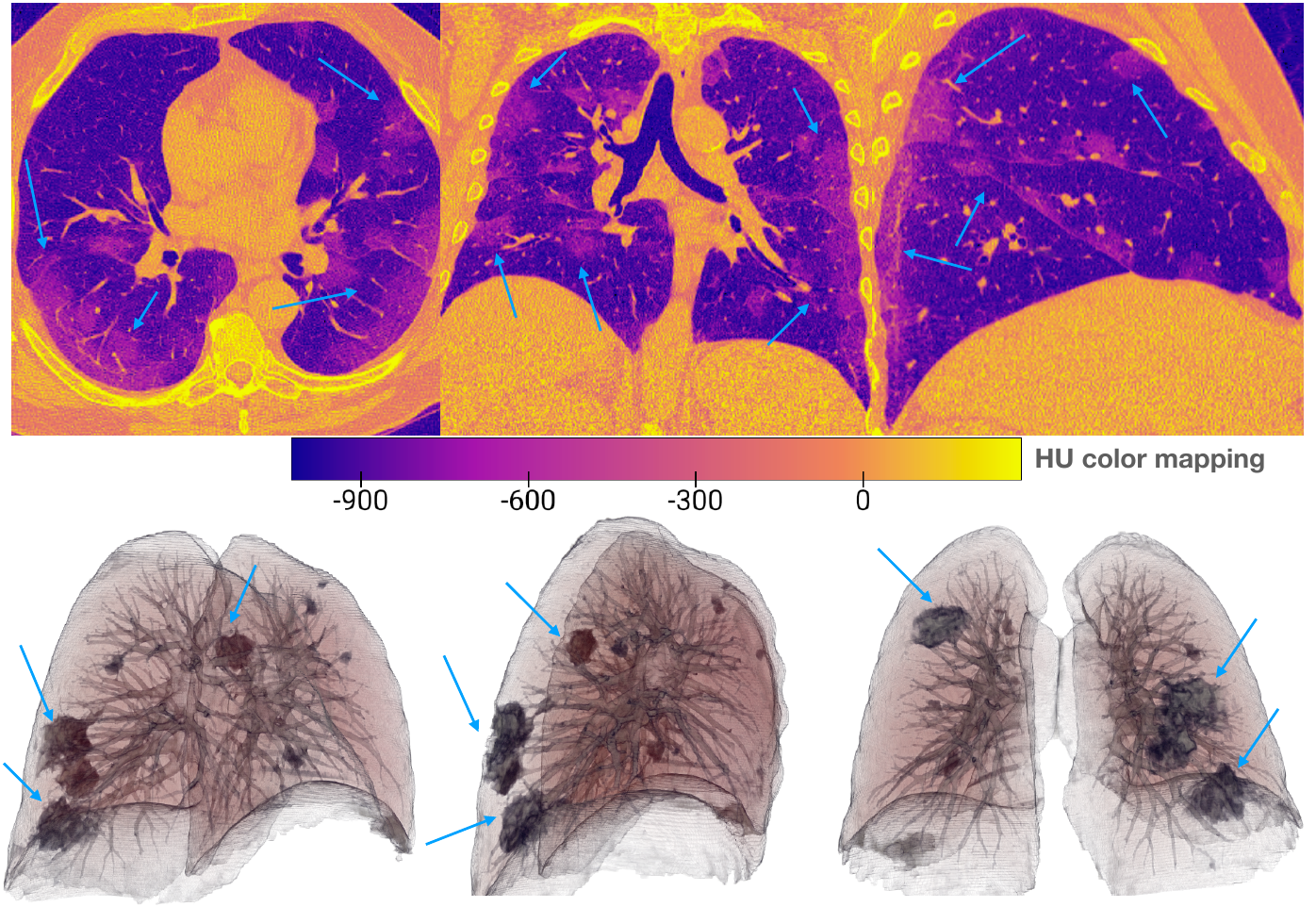}
    \caption{\cv lesions in CT slices and volume renderings of the lung: (top) axial, coronal, and sagittal CT slices display ground-glass opacities associated with \cv lesions correspond to light magenta areas inside the lungs; (bottom) volume rendering from different angles reveal ground-glass opacities associated with \cv lesions as darker regions inside the lungs}
    \label{fig:classes}
\end{figure*}

During the diagnosis process, radiologists search for characteristic patterns of lung lesions associated with Ground-Glass Opacity (GGO)\cite{Kwee2020ChestKnow}. GGO appears as a gray or hazy region of increased attenuation in the lung. In the case of \cv infection, GGO exhibits specific features such as its location (peripheral or bilateral), shape (rounded), and appearance (multifocal or closer to opaque lung tissues known as consolidations). For instance, on top of Figure\ref{fig:classes}, the GGOs corresponding to \cv are depicted as light magenta areas within the lungs.

A limitation of CT-based diagnosis is that it consumes around 20 minutes of a radiologist's time. To automate this task, machine learning solutions have emerged, such as Deep Learning (DL) architectures based on Convolutional Neural Networks (CNNs) that use as input images the slices of a CT~\cite{Roberts2021,ALHASAN2021101933}. A slice-based architecture is a common solution since it reflects the local approach radiologists take when looking at individual slices of a CT. Numerous proposals have been presented in the literature, and we refer the reader to a systematic literature review of the most relevant systems~\cite{HASSAN2022105123}. 

Despite the vast amount of research conducted on this topic, the scientific community is continuously striving to improve the accuracy and effectiveness of existing diagnostic models.
In this work, we propose an alternative that follows a global approach of looking at images of the entire lung. We propose to generate volume-rendering images that use transparency to capture the inner structures of the lung, taken from different angles to overcome possible occlusions. The bottom of Figure~\ref{fig:classes} displays volume rendering images produced from different angles, in which GGOs are associated with darker regions in the lung.

Our approach, COVID-VR, is based on a 3D Volume Rendering classification architecture consisting of
a pipeline composed of three main modules.
The first module receives as input a chest CT and performs a segmentation step that removes material outside the lung. The second module receives a segmented CT and generates volume rendering of the lung from different angles. The third module comprises a ResNet architecture~\cite{RESNET} that receives volume rendering images and outputs a classification in two or three classes.

We tested COVID-VR using the \PUBDATASET~\cite{Afshar2021}\footnote{\url{https://github.com/ShahinSHH/COVID-CT-MD}} public dataset and a private dataset obtained from partner hospitals in Porto Alegre (Brazil). We present classification results comparing our architecture against three competing strategies, showing that our approach is competitive and can potentially be an alternative for \cv classification from CTs.
\section{Related Work}
\label{sec:related-work}
This section summarizes related works relevant to our proposal. First, we review deep learning-based models described for COVID-19 diagnosis and the main candidates for comparing the results of our work. Next, we review the use of volume rendering for a better understanding of COVID-19.


\subsection{DL-based models for COVID-19 diagnosis from CTs}

%
%
DL techniques are widely utilized for the classification of CT images. To develop an effective classification system, it is crucial to comprehend the patterns associated with each class. Guidelines for classifying \cv based on CT images can be found in Radiology standards. The Radiology Society of North America (RSNA) and the British Thoracic Imaging Society (BSTI) have provided similar classification schemes for COVID-19, consisting of four categories.
The RSNA categories are \cv typical, indeterminate, atypical, and negative for pneumonia. The typical classification corresponds to patients with lesions associated with ground-glass opacities, such as those seen in Figure~\ref{fig:classes}. The indeterminate classification suggests the absence of typical features, plus multifocal, diffuse, perihilar, nonperipheral, nonrounded, or unilateral ground-glass opacity with or without consolidation. Atypical classification denotes the absence of the previous classifications' features and the presence of isolated consolidation without ground-glass opacities or discrete small nodules. The negative classification indicates no features of pneumonia.
Even though both standards describe four classes, the first models to appear in the literature were either binary (COVID-19 or non-COVID-19) or ternary (COVID-19, normal, or Community-Acquired Pneumonia (CAP)). The normal and CAP classes are similar to the negative and indeterminate classes of RSNA. 

One way to navigate the extensive literature on the topic is to read the survey papers that summarize the proposed systems. 
The survey paper by Hassan et al.~\cite{HASSAN2022105123} is one of the most complete references on the topic, offering a categorization of CT-based diagnosis methods (classification, segmentation, and detection), frameworks and relevant findings and a list of open-source datasets. In total, they summarize 114 studies. 


We summarize below the related work to our proposal that uses CT images as inputs to their classification models~\cite{Jin2020,Wang2020,Li2020,He2020,Wang2020b,Gozes2020,AMYAR2020104037,IBRAHIM2021104348,SERTE2021104306}.
DeCoVNet~\cite{Wang2020} uses a weakly-supervised 3D deep convolutional network to predict the probability of COVID-19 following a binary classification approach. In the first step, 
the authors used a U-NET to segment the lung in the volume and create a 3D binary mask. The CT and its 3D lung mask are sent to DeCoVNet, which consists of three stages: 3D convolution, 3D residual blocks, and a progressive classifier. 
Ternary models soon followed binary models to allow the separation of COVID-19 from normal cases and other types of CAPs. An example is COVNet~\cite{Li2020}, which uses a 3D deep learning framework that takes 3D slices as input, generates features, combines them using a max-pooling operation, and finally generates a probability score for each class. Like DeCoVNet, COVNet also relies on a pre-processing step that uses a U-NET to create a segmented mask for the lung. Amyar et al.~\cite{AMYAR2020104037} proposes a ternary model that uses a shared encoder for the classification and lesion segmentation tasks without requiring labeled segmentation data. Deep-chest~\cite{IBRAHIM2021104348} uses a VGG-CNN model that supports classification in four classes: COVID-19, pneumonia, lung cancer, and normal cases. Serte and Demirel~\cite{SERTE2021104306} proposed a binary 3D classification model that fuses image=level predictions to classify CT volumes. Silveira et al.~\cite{DASILVEIRA2023103775} compute omnidirectional images from the center of the lungs and use them as inputs to the DL classification model. Our approach is similar in the sense that it uses volume rendering images of the lungs, but from an external point of view. 

The ICASSP 2021 competition brought an opportunity to compare methods by providing a public dataset of CTs (COVID-CT-MD~\cite{Afshar2021}) and a contest of performing classification in three possible classes: normal, COVID-19, or CAP. The six-best solutions were presented at the ICASSP conference~\cite{Chaudhary2021,Pratyush2021,Zaifeng2021,Xue2021,Bougourzi2021,Li2021}.
The first place (team TheSaviours~\cite{Chaudhary2021}) has an accuracy of 90\%. They describe a two-stage CNN model for detecting COVID-19 and Community-Acquired Pneumonia (CAP). The first stage is designed to detect infections (COVID-19 or CAP). The second stage performs the classification into three classes (CAP, COVID-19, and Normal). 
The second place 
(team IITDelhi~\cite{Pratyush2021}) has an accuracy of 88.89\%. Their method follows a three-level approach. In the first level, they use a slice-level classifier that performs feature extraction from all the slices of the CT to learn different sizes of infection. The next level performs a patient-level classifier, using four classifiers to distinguish between infected and normal slices. The last level uses an ensemble-learning that combines the scores of the previous level classifiers.
The third place (team LLSCP ~\cite{Zaifeng2021}) has an accuracy of 87.78\%.  They use a multi-stage progressive learning approach composed of a 3D Resnet module, an ensemble binary classifier, and a final combining stage. The remaining entries achieved 85.56\%~\cite{Xue2021},  81.11\%~\cite{Bougourzi2021} and 80\%~\cite{Li2021}.

For comparison purposes, we chose TheSaviours~\cite{Chaudhary2021} (ICASSP 2021 winner), DeCoVNet~\cite{Wang2020}, and COVNet~\cite{Li2020} approaches due to their outstanding performance and the public availability of code or pre-trained models.
\subsection{Volume Rendering Applications in \cv}
Volume Rendering combines opacity mapping and lighting effects values through multiple rendering techniques \cite{arens2010survey, jonsson2014survey, kniss2002multidimensional}.
It is widely used in medical image visualization \cite{zhang2011volume} to analyze internal spatial relationships between structures, but few studies report its use in the \cv context. 

Tang et al.~\cite{Tang2020SevereCT} report one of the first efforts for visualizing \cv pneumonia, displaying lung lesions in a color coronal image and a tridimensional volume rendering of the lungs, bronchus, and trachea from a 54 years old patient. 
Li et al.~\cite{li2020multiscale} show the advantage of 3D volume rendering to detect the extent of small pulmonary vessel microangiopathy and alveolar damage from an autopsy of a \cv patient. COVID-view~\cite{Jadhav2021} describes a system for COVID-19 diagnosis that combines a classification system with explainable visualizations of activations over 2D slices and volume rendering of the lungs. They use Maximum Intensity Projection (MIP) to create projections of volumes that help identify vascular structures. In our work, we use Volume Rendering instead of MIP, and remains an issue for future work on how MIP would work in our system. COVID-view user interface allows users to modify the transfer functions used for volume rendering by choosing from a preset transfer function or by creating their own. They also use a coronal clipping tool that allows better inspection of the inner structures inside the lungs. However, their proposed classification model does not rely on volume rendering images but uses 2D slices. Similarly, COVIR~\cite{AMARA202211} proposes a classification model based on 2D slices and a virtual reality platform to explore 3D reconstructed lungs and segmented infected lesions caused by COVID-19. The system relies on software such as Blender to fix problems with the models and Unity to produce the final renderings and support the interaction through VR devices.

\section{CT Datasets}
\label{sec:datasets}
This study used \cv-related CT datasets from three different sources. The first source is the \textit{Public} dataset of CT images prepared for the ICASSP 2021 competition.
The other two sources are from partner hospitals in Brazil that provided CT scans for COVID-19 and non-COVID-19 cases. These datasets are referred to as \textit{Private} datasets. CTs for all datasets are given in the DICOM format. This section describes the details of the creation and statistics of these datasets.



\subsection{COVID-CT-MD Public Dataset}
\label{sec:datasets-public}
The COVID-CT-MD public dataset~\cite{Afshar2021} was released for the SPGC-ICASSP competition and adopted for model development in other works (\eg \cite{Chaudhary2021,Pratyush2021}), thus allowing the comparison of our approach against state-of-the-art methods. The dataset comprises 307 CT scans with patient-level annotations divided into three classes: confirmed positive COVID-19 cases, normal cases, and CAP cases. The COVID-19 cases were collected from February to April 2020, while CAP and normal cases were collected from April 2018 to December 2019 and January 2019 to May 2020, respectively. Although slice-level annotations are available for some patients, we did not explore them in our approach. 
The labeled CT scans used for model training and validation are based on a stratified random split: 30\% of these CT scans are randomly selected as the validation set, and the remaining are used as the train set. Patients are adults recruited in the Babak Imaging Center, Iran, and exam labeling was conducted by experienced radiologists~\cite{Afshar2021}. Table \ref{tab:publicDataset} shows the number of CT scans per class. 


In addition to the train/validation set, the SPGC-ICCASP competition released the SPGC-COVID Test Set \cite{heidarian2021robust}, with four independent test sets for model evaluation. Three sets were used to calculate the competition results and are applied for performance comparison of our work with previous approaches. Each test dataset contains 30 CT scans, as described by Heidarian et al. \cite{heidarian2021robust}. The first set (Test Set 1) comprises \cv and Normal cases (15 and 15, respectively), obtained from the same image center as the train/validation set. The second set (Test Set 2) contains the three classes, \cv, Normal, and CAP, with 10 cases for each one, obtained from another imaging center (Tehran Heart Center, Iran) using different scanner and scanner parameters. The third set (Test Set 3) contains the three classes, \cv, Normal, and CAP, with ten samples for each one collected in the same imaging center as Test Set 1. Test Set 2 differs from the others because it includes patients with a history of cardiovascular diseases and surgeries. The distribution of samples per class for the SPGC-COVID Test Sets is given in Table \ref{tab:publicDataset}.



\subsection{Private Datasets from Partner Hospitals }
\label{sec:datasets-private}
The private datasets were retrospectively obtained from patients admitted to two Brazilian hospitals in Porto Alegre: a private institution, Hospital Moinhos de Vento (HMV), and a public institution, Hospital de Clínicas de Porto Alegre (HCPA). The Research Ethics Committee in Brazil approved the study of the participating hospitals, and informed consent was waived due to the study retrospective nature\footnote{HCPA approval - CAAE: 32314720.8.0000.5327, HMV approval - CAAE: 32314720.8.3001.5330}. All data were anonymized by providers to ensure patient privacy. The HMV dataset contains 284 CT scans from patients admitted between March and May 2020, whereas the HCPA dataset comprises 105 CT scans collected from March to June 2020. Both datasets have patient-level annotations by expert radiologists, following the RSNA standard~\cite{simpson2020radiological}. As previously explained (Section~\ref{sec:related-work}), the RSNA standard classifies images into four classes: Typical appearance, Indeterminate appearance, Atypical appearance, and Negative for pneumonia. The first category corresponds to exams showing CT features frequently seen in patients with COVID-19 pneumonia (\eg bilateral, peripheral, and multifocal ground-glass opacities), thus representing our class of interest (\ie positive). The distribution of CT scans among classes is shown in Table~\ref{tab:privateDataset}. 


\begin{table}[!tb]
\centering
\caption{COVID-CT-MD (Public dataset): number of CTs and distribution by class.}
\begin{tabular}{cllllll}
\toprule
\textbf{Class}               & \multicolumn{3}{c}{\textbf{\begin{tabular}[c]{@{}c@{}}Train/Validation \\ (307)\end{tabular}}} 
& \multicolumn{3}{c}{\textbf{\begin{tabular}[c]{@{}c@{}}Test\\ (90)\end{tabular}}} \\
\hline
                             & F                             & M                              & Total                         & F                         & M                         & Total                    \\
\hline
\multicolumn{1}{l}{COVID-19} & \multicolumn{1}{r}{63}        & \multicolumn{1}{r}{108}        & \multicolumn{1}{r}{171}       & \multicolumn{1}{r}{9}     & \multicolumn{1}{r}{26}    & \multicolumn{1}{r}{35}   \\
\multicolumn{1}{l}{Normal}   & \multicolumn{1}{r}{36}        & \multicolumn{1}{r}{40}         & \multicolumn{1}{r}{76}        & \multicolumn{1}{r}{15}    & \multicolumn{1}{r}{20}    & \multicolumn{1}{r}{35}   \\
\multicolumn{1}{l}{CAP}      & \multicolumn{1}{r}{26}        & \multicolumn{1}{r}{34}         & \multicolumn{1}{r}{60}        & \multicolumn{1}{r}{7}     & \multicolumn{1}{r}{13}    & \multicolumn{1}{r}{20}   \\
\bottomrule               
\end{tabular}
\label{tab:publicDataset}
\end{table}

\begin{table}[!tb]
\centering
\caption{Private dataset: number of CTs and distribution by class.}
\begin{tabular}{lcclccc}
\toprule
\textbf{Class} & \multicolumn{3}{c}{\textbf{\begin{tabular}[c]{@{}c@{}}HMV\\(289)\end{tabular}}} & 
\multicolumn{3}{c}{\textbf{\begin{tabular}[c]{@{}c@{}}HCPA\\ (105)\end{tabular}}} \\
\hline
& \multicolumn{1}{c}{F} & \multicolumn{1}{c}{M} & \multicolumn{1}{c}{Tot.} & \multicolumn{1}{c}{F} & \multicolumn{1}{c}{M} & \multicolumn{1}{c}{Tot.} \\
\hline
Typical & \multicolumn{1}{c}{34} & \multicolumn{1}{c}{58} & \multicolumn{1}{c}{92} & \multicolumn{1}{c}{13} & \multicolumn{1}{c}{17} & \multicolumn{1}{c}{30}\\
Negative & \multicolumn{1}{c}{54} & \multicolumn{1}{c}{38} & \multicolumn{1}{c}{92} & \multicolumn{1}{c}{11} & \multicolumn{1}{c}{4} & \multicolumn{1}{c}{15}\\
Indeterm. & \multicolumn{1}{c}{32} & \multicolumn{1}{c}{35} & \multicolumn{1}{c}{67} & \multicolumn{1}{c}{16} & \multicolumn{1}{c}{14} & \multicolumn{1}{c}{30}\\
Atypical & \multicolumn{1}{c}{16} & \multicolumn{1}{c}{17} & \multicolumn{1}{c}{33} & \multicolumn{1}{c}{13} & \multicolumn{1}{c}{17} & \multicolumn{1}{c}{30}\\
\bottomrule
\end{tabular}
\label{tab:privateDataset}
\end{table}

\section{The COVID-VR Classification Architecture}
\label{sec:covid-vr}

\begin{figure*}[!tb]
    \centering
    \includegraphics[width=0.99\linewidth]{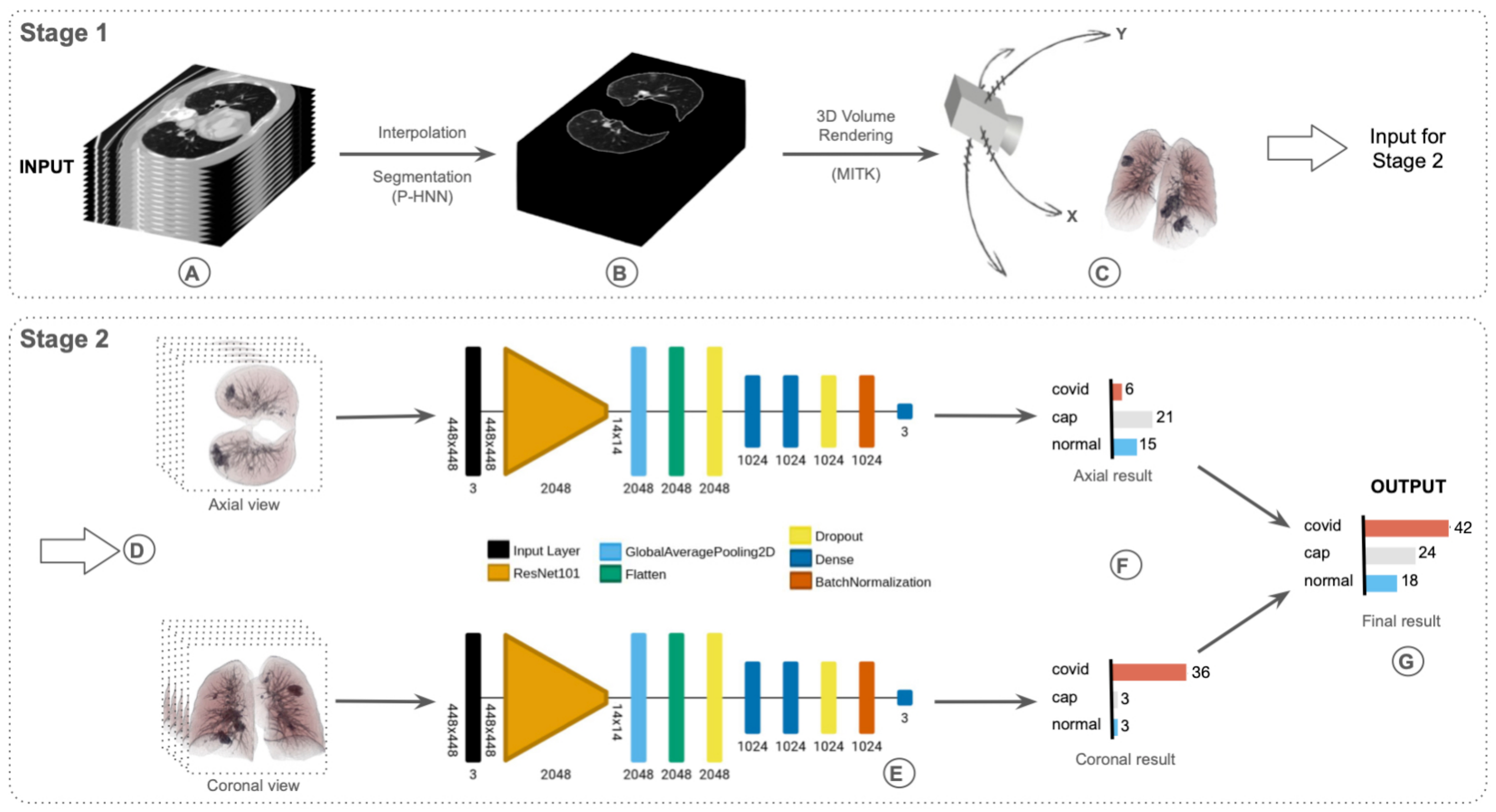}
     \caption{The pipeline of COVID-VR has two stages. Stage 1 prepares data to obtain the input images for model development. In Stage 2, DL models are trained to distinguish among the classes of interest, and their outputs are combined to obtain a final patient-level classification.}
    \label{fig:pipeline}
\end{figure*}

In this section, we describe COVID-VR, an end-to-end pipeline for the analysis of CT images, from lung segmentation to patient-level classification associated with the presence of COVID-19 based on volume rendering images (Figure~\ref{fig:pipeline}).


\subsection{Lung Segmentation}
The first step is to convert the input CT images (Figure~\ref{fig:pipeline}A) from the DICOM format to the NIfTI (Neuroimaging Informatics Technology Initiative) format. We use the ALTIS system~\cite{Sousa2019} to obtain isometric volumes across patients, which requires interpolating and resizing images with a slice spacing of 1mm in all dimensions. 
Several methods for lung segmentation were considered, such as UBC \cite{lensink2020segmentation}, Lungmask \cite{hofmanninger2020lungmask}, and P-HNN~\cite{PHNN}.
P-HNN was selected after recommendations by radiologists who compared the results of these methods. Lung segmentation is performed with the pre-trained model of P-HNN, using a probability mask with a threshold of 75\% or above to consider a voxel as part of the lung (Figure~\ref{fig:pipeline}B). 

It is important to observe that our proposal does not require highly-accurate segmentation results, but enough to discard additional volume around the lungs. Therefore, we did not perform additional testing to check the accuracy of the segmentation. The images generated with volume rendering rely on transfer functions that associate different opacity values that can compensate for small errors in the segmentation algorithm. 



\subsection{Generating Volume Rendering Images}
\label{sec:vr}

This step generates volume rendering images from the lung segmented in the previous stage (Figure~\ref{fig:pipeline}C). There are two main problems that we need to address to configure the volume rendering algorithm. The first one is the specification of a transfer function (TF) that reveals the internal structures of the lung and lesions associated with \cv. The second problem is choosing the camera positions to generate the volume rendering images. Ideally, we want to generate images from different points of view that better capture the lungs and lesions. Both issues are described in this section. 
%
%

\begin{figure*}[!tb]
\centering
     \includegraphics[width=0.9\linewidth]{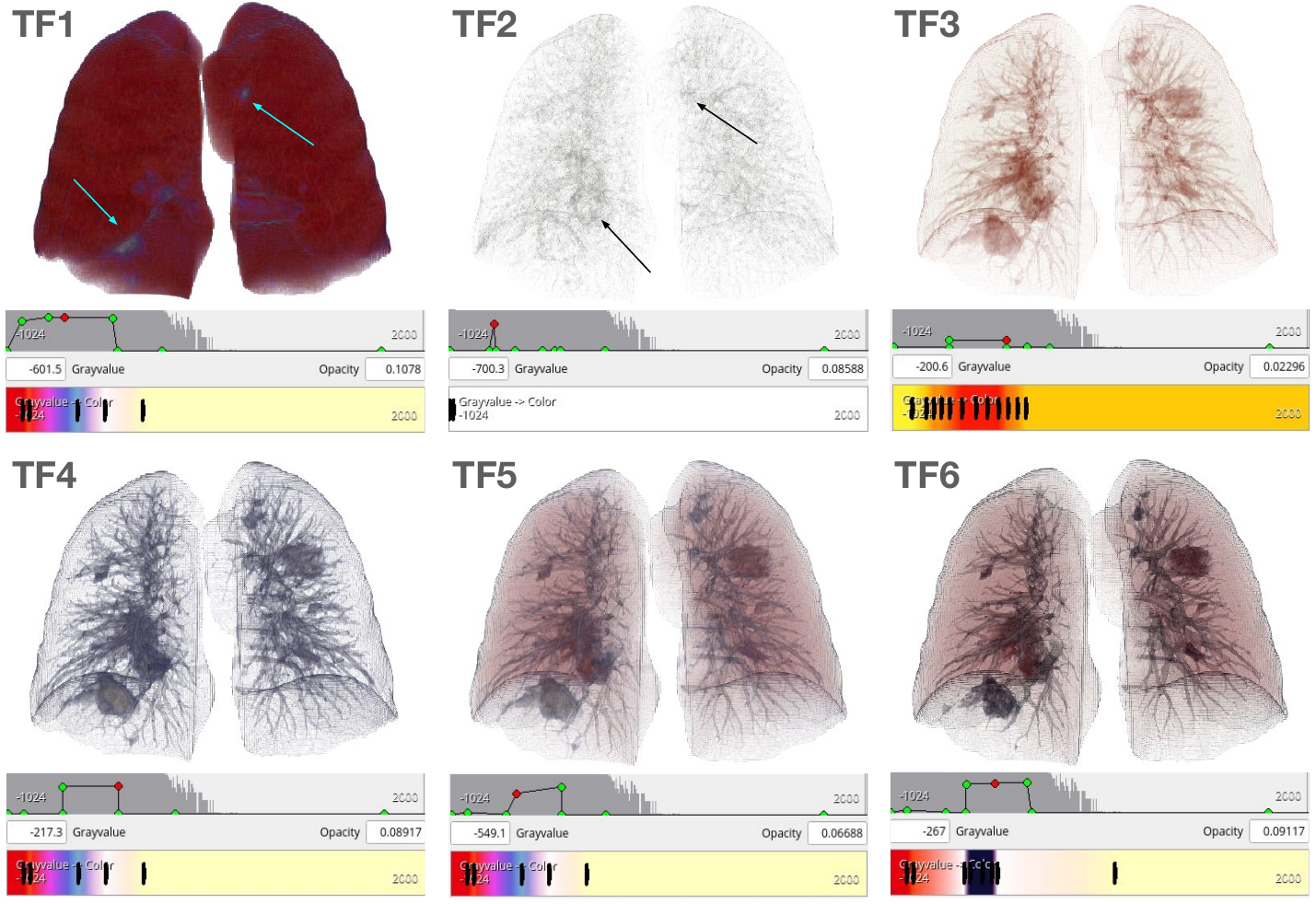}

     
     
\caption{Volume rendering of the lung using customized transfer functions. TF1 shows the boundary of the lungs but misses internal structures. TF2-TF4 shows internal structures but has similar colors for both GGOs (lesions) and the lung. TF5-TF6 better separates lungs and lesions.} 
\label{fig:tfs}
\end{figure*}

\subsubsection{Transfer Function Selection}
\label{sec:vr-tfs}

Volume rendering relies on a transfer function (TF) specification that defines a mapping from input values to color, transparency, and opacity. The input values in the segmented lungs are expressed in Hounsfield units (HU), a dimensionless scale obtained from a linear transformation of the measured absorption/attenuation of the X-ray beam. Although there is no universal standard for defining HU intervals for the lungs and GGOs, related works~\cite{tang2020severity, lu2021quantitative} report values of around -700 HU for the lung and in the interval from -700 HU to -300 HU for the GGOs. We use this information to customize TFs. 

We use the MITK framework~\cite{wolf2004medical} to conduct tests with different TFs and to create the volume rendering images from the 3D volume of the segmented lungs.
MITK has a user interface that allows for building customized transfer functions. 
In our tests, we explored more than 60 different TFs. Figure~\ref{fig:tfs} shows volume rendering images using six different TFs (TF1 to TF6) defined experimentally. 
Although they were all applied to the same CT exam, the rendered image differs significantly. TF1 highlights the outer layer of the lungs but misses internal structures.
TF2 to TF6 reveal internal lesions in different ways.
TF2 is defined within the lower range of values in the lung and ground-glass opacity (\ie [-700, -300]), mapping only HU values in this interval. This function conserves a particular spatial distribution of the ground-glass opacity present in the lung but misses details on the lung texture and regions without any lesions. TF3 aims to replicate the behavior of 2D CTs that vary in a single color scale (usually gray-scale), mapping original values in the range of [-750, -200] with variations applied to the brightness. TF3 manages to preserve the lung as a three-dimensional image without losing critical information on the internal lesions caused by COVID-19, such as ground-glass opacity. TF4 to TF6 are obtained by including more colors to help differentiate the lung textures and features of the different classes. 
 TF4 shows a thin delineation of the lung inner layer and a prominent highlighting of the bronchi, which in the presence of \cv may show changes such as bronchial wall thickening \cite{ye2020chest}. For TF5 and TF6, we observe a more precise differentiation of ground-glass opacities from other characteristics, which may be interesting for the classification model as this feature is the most common imaging finding in \cv patients \cite{ye2020chest}. Thus, by exploring a wide range of mapping combinations for color and opacity, each TF can highlight different regions or features of interest when applied to CT scans, such as external surfaces of the lungs as shown in TF1, arteries as shown in TF3, and ground-glass opacity as in TF5 and TF6.

\subsubsection{Choosing the Viewing Camera Positions}
\label{choosing-view-positions}
Volume rendering images need to capture the lungs and lesions. An important issue is the placement of the viewing camera in a location outside the 3D model of the lung. We explored placing camera positions looking at the main planes (axial, coronal, and sagittal). We  discarded the sagittal plane because it renders one lung in front of the other. Due to transparency, the volume rendering of overlapping lungs might lead to combined patterns that might confuse the classification model. 

For axial and coronal planes, we rendered images starting from a position orthogonal to each plane and rotated the camera position along camera increments to capture the lungs from slightly different angles. We tested different angle increments and relied on the results from the classification model to fine-tune our choices. We captured images at increments of $\pm$1.2° degrees in the Y and X-axis until reaching a maximum of 12.0° degrees. 
There are 21 images on each of the Y and X-axis, leading to 42 different images per view (\ie coronal or axial), for a total of 84 images per dataset. 

\begin{figure*}[!tb]
    \centering
    \includegraphics[width=0.98\linewidth]{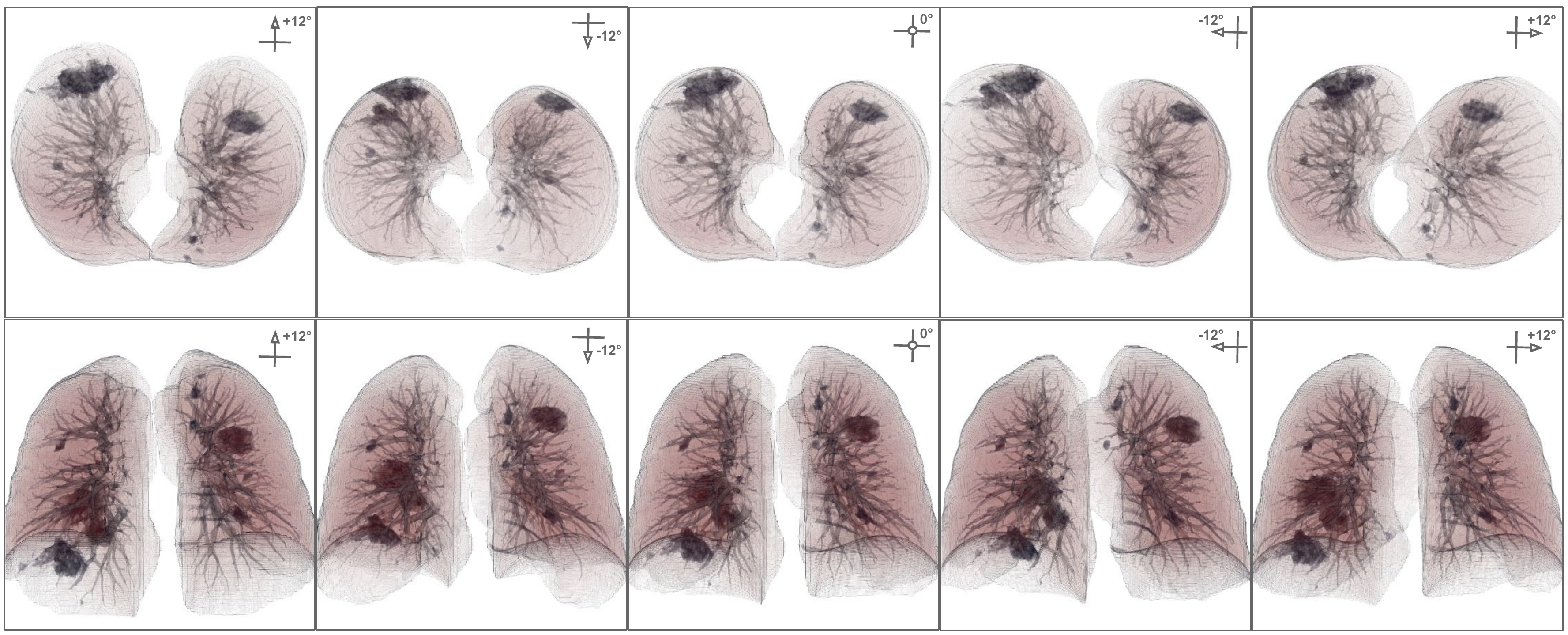}
    \caption{Volume rendering images used as inputs to the classification model. The first row represents images extracted for the axial plane. The third image in this row shows the (0,0) position in which the camera points to the axial view. From this position, we capture images every 1.2° towards the horizontal and vertical axis directions, going from -12° to +12° on each axis. The second row shows the same process for the coronal plane.}
    \label{fig:lungs_2axis}
\end{figure*}

The resolution of each image was defined to be 448$\times$448px~(Figure~\ref{fig:pipeline}D) due to the constraints posed by the classification model network. Examples of the resulting images from different angles are shown in Figure~\ref{fig:lungs_2axis}.

\subsection{Classification Models}
\label{sec:ClassificationModel}
COVID-VR is composed of deep neural networks (DNN) models developed using the Tensorflow framework (Figure~\ref{fig:pipeline}E). In this section, we detail the architecture and training of the models, as well as their evaluation.

\subsubsection{Model Architecture and Training}
We use transfer learning to improve models for COVID-19 classification.
We used a pre-trained Convolution Neural Network (CNN) model 
as the backbone for the architecture, such as the VGG~\cite{VGG}, Resnet~\cite{RESNET}, DenseNet~\cite{DENSENET}, or EfficientNet~\cite{EFFICIENTNET}.
Figure~\ref{fig:pipeline}E shows the Resnet101 as the backbone network, represented in orange. We add a sequence of deep learning modules to this backbone to help model training. For example, following the backbone, we add a global average pooling layer, a 20\% dropout layer to avoid overfitting, two fully-connected dense layers, a new 20\% dropout layer, and a batch normalization layer. The final module is a dense layer with a Sigmoid activation function for the binary classification task and a Softmax activation function for the ternary classification task. The selection of which network architecture we would use was performed empirically through several preliminary experiments. For model compilation, we use the Adam optimizer with a learning rate of $2e-5$, with a binary (categorical) Cross-entropy loss function for binary (ternary) models.

Model training and validation were conducted using stratified 5-fold cross-validation (CV) for the private datasets. We follow the competition orientation for the public dataset and use a random 70\%/30\% split from the COVID-CT-MD dataset \cite{Afshar2021} to generate the train and validation sets, respecting original class distributions among splits. The model is further evaluated with three independent test sets provided by the SPGC-ICASSP competition \cite{heidarian2021robust}, as previously described (Section~\ref{sec:datasets-public}). Additionally, to increase the size and variety of the dataset at the training step and reduce the chances of overfitting, we use data augmentation methods in the training partition, like rotation up to 15°, zoom ($\pm$5\%), rescaling of 1/255, and shift for width and height images up to 10\%. 

We train an individual DNN for each defined plane, either axial or coronal, based on the image views extracted from it. Moreover, despite the use of patient-level annotations for the supervised learning task, the output of our DNNs is a classification per image view. Thus, the batch of image views obtained for a given CT scan (\eg 42 axial views or 42 coronal views) is classified by the corresponding network, specialized in either the axial or coronal view, generating a class label for each image analyzed. We obtain a distribution of class votes for each image batch that passes a consensus-extraction module to generate a patient-level prediction.

Figure~\ref{fig:pipeline}F illustrates the classification approach. It is defined based on two submodels, each predicting the \cv diagnosis using images obtained from a specific view of the reconstructed 3D volume, axial or coronal. We generate a patient-level consensus distribution based on the distribution of class votes received from each submodel by summing up the number of votes per class. For example, in the ternary classification task shown in Figure~\ref{fig:pipeline}F, the 42 images generated from the axial view are classified by the corresponding submodel with the following distribution: 6 votes for class \cv, 21 votes for class CAP, and  15 votes for class Normal. On the other hand, the submodel trained for coronal view images assigns 36 votes for class \cv, three votes for class CAP, and three votes for class Normal. The ensemble-based solution generated from the sum of submodels votes results in 42 images classified as \cv, 24 images classified as CAP, and 18 images classified as Normal. Figure~\ref{fig:pipeline}G shows the final model used to predict a single label for the input CT scan and the class with the maximum number of votes according to both views. In this example, the CT would be classified as \cv by our approach.


\subsubsection{Model Evaluation}
We use the following evaluation metrics to assess the performance of our \cv models: 
accuracy (Acc), sensitivity (Sens, also called recall), specificity (Spec), precision (Prec), F1-score (\ie the harmonic mean between recall and precision), and the area under the Receiver Operating Characteristic (ROC) Curve (AUC score). Given its multiclass nature in the ternary model, we adopted the micro-and macro-average for all metrics except accuracy.

\section{Results}
\label{sec:results}


This section reports the evaluation of the models proposed in COVID-VR. We first compare various backbone networks and transfer functions using the public dataset (COVID-CT-MD). We use the insights from these results in subsequent experiments. Next, we compare the classification performance of the models with state-of-the-art approaches for detecting COVID-19-related pneumonia in the public dataset. Finally, we discuss the performance of the models developed from the private datasets and compare both ternary and binary classification models. Since there is a class imbalance in the datasets, we focus our discussion on the micro-average metric for the ternary models. Full results are provided in Table~\ref{tab:all-results-micro} and Table~\ref{tab:all-results-macro}.

\subsection{Selection of the Backbone Network}
We tested several CNN architectures as the backbone network to perform CT scan classification with the models described in Section\ref{sec:ClassificationModel}, including ResNet~\cite{RESNET}, DenseNet~\cite{DENSENET}, VGG~\cite{VGG}, and EfficientNet~\cite{EFFICIENTNET} families. 
We explored different depths for each architecture (\ie ResNet50, ResNet101, DenseNet121, DenseNet201, EfficientNet-B0, EfficientNet-B1, EfficientNet-B6, VGG16, and VGG19) with fixed training and validation sets contained in the COVID-CT-MD dataset. A comparison of the classification performance of the better models for each network family is given in Table \ref{tab:ComparisonArchitecturesInSPGC}. We use a fixed transfer function (TF6) to generate volume rendering images in all experiments. The choice of the transfer function is explained in the next section. The performance results correspond to  the ternary classification models using the train and validation sets from the public dataset (\ie COVID-CT-MD dataset). We show the performance for the \cv class and the overall performance for the three classes. F1 and AUC scores are obtained with the micro-average. Additional results per class can be found in Table~\ref{tab:all-results-micro}.

\begin{table}[!tb]
\footnotesize
\centering
 \caption{Comparison of distinct backbone network architectures in COVID-VR. Model training and validation use the train and validation sets from the COVID-CT-MD public dataset for the ternary classification task (\cv vs. CAP vs. Normal).} 
\label{tab:ComparisonArchitecturesInSPGC}
\begin{tabular}{@{}clcccc@{}}
\toprule
\multicolumn{1}{l}{} & Metrics & VGG16 & DenseNet121 & EfficientNet-B2 & ResNet101 \\ \midrule
\multirow{3}{*}{Overall} & Acc & 88.8\% & 87.8\% & 86.7\% & \cellcolor[HTML]{FFFF00}{\textbf{90.8\%}} \\
 & F1 & 88.6\% & 87.7\% & 86.7\% & \cellcolor[HTML]{FFFF00}{\textbf{90.8\%}} \\
 & AUC & 95.6 & \cellcolor[HTML]{FFFF00}{\textbf{96.5}} & 95.1 & 95.4 \\\midrule
\multirow{4}{*}{COVID-19} & Sens & \cellcolor[HTML]{FFFF00}{\textbf{89.1\%}} & 85.5\% & 83.6\% & \cellcolor[HTML]{FFFF00}{\textbf{89.1\%}} \\
 & Spec & 88.4\% & \cellcolor[HTML]{FFFF00}{\textbf{93.0\%}} & 90.7\% & \cellcolor[HTML]{FFFF00}{\textbf{93.0\%}} \\
 & Prec & 90.7\% & 94.0\% & 92.0\% & \cellcolor[HTML]{FFFF00}{\textbf{94.2\%}} \\
 & F1 & 89.9\% & 89.5\% & 87.6\% & \cellcolor[HTML]{FFFF00}{\textbf{91.6\%}} \\ \midrule 
\end{tabular}%
\end{table}

Our ternary model achieved the best results using the ResNet101 network as its backbone. The validation accuracy and F1-score were the highest among the best models for each network family. Both accuracy and F1-score for the overall classification in a ternary approach were 90.8\%, approximately two percentage points above VGG16, which presented the second-best performance (Table~\ref{tab:ComparisonArchitecturesInSPGC}). Regarding AUC score, DenseNet121 achieved the highest mark of 96.5 compared to 95.4 for ResNet101. Nonetheless, observing the classification results for the COVID-19 class (Table~\ref{tab:ComparisonArchitecturesInSPGC}), we note that ResNet101 showed the best performance for all metrics. Moreover, it achieved the best performance for CAP cases and competitive performance for Normal cases (see Table~\ref{tab:all-results-micro}). Thus, we chose ResNet101 as the backbone network for COVID-VR, using it for all the subsequent experiments.

\subsection{Selection of the Transfer Function}
We compared the transfer functions discussed in Section~\ref{sec:vr-tfs} to decide which one we would use to generate volume rendering images. We used the results of the ternary classification model for this purpose, with the train and validation sets from the public dataset. Table~\ref{tab:ComparisonTFsInSPGC} summarizes the results. TF4 and TF5 were omitted due to similar performance obtained with TF3 and TF2, respectively. TF6 presented the best accuracy and overall F1 score performance. Although TF6 does not have the highest sensitivity, it had a balanced performance in detecting \cv cases. In addition, TF6 achieved the best sensitivity for CAP and normal cases.  

\begin{figure}[!tb]
    \centering
     \includegraphics[width=1.0\linewidth]{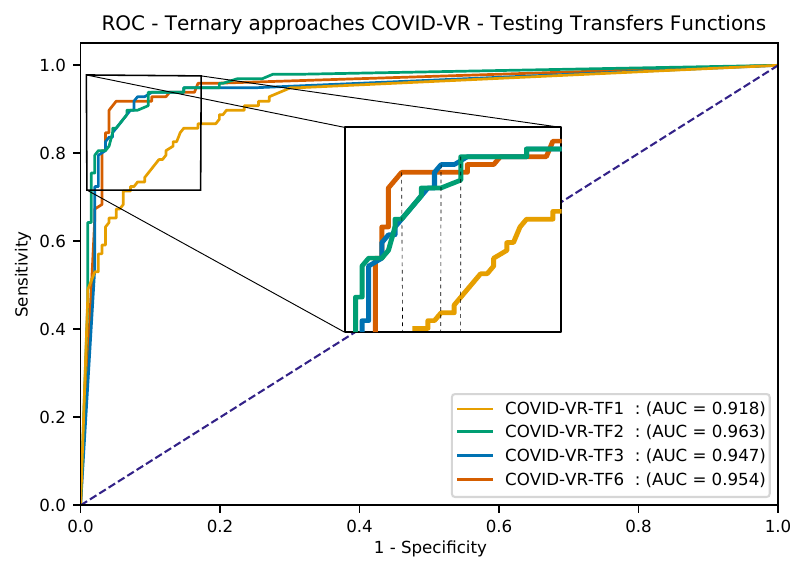}
    \caption{ROC curves for transfer functions TF1, TF2, TF3, and TF6}
    \label{fig:ROC-comparison-spgc-tfs}
\end{figure}

Figure \ref{fig:ROC-comparison-spgc-tfs} compares the ROC curves obtained for distinct TFs. Although there is a small between the AUC score for TF6 and TF2, the ROC curve for TF6 achieves the highest sensitivity for a 10\% false-positive rate. Therefore, we chose TF6 as the standard transfer function. Besides the visual detection of different features highlighted by each TF, this test proved relevant, as switching the TF while keeping the same model architecture results in considerable accuracy variation.

\begin{table}[!tb]
\footnotesize
\caption{Comparison of distinct transfer functions in COVID-VR.
Model training and validation use the train and validation sets from the COVID-CT-MD public dataset for the ternary classification task (\cv vs. CAP vs. Normal).}
    \label{tab:ComparisonTFsInSPGC}
\centering
\begin{tabular}{clcccc}
\hline
\multicolumn{1}{l}{} & Metrics & TF1 & TF2 & TF3 & TF6 \\ \hline

\multirow{3}{*}{Overall} & Acc & 79.6\% & 85.7\% & 87.8\% & \cellcolor[HTML]{FFFF00}{\textbf{90.8\%}} \\
 & F1 & 78.7\% & 84.5\% & 87.6\% & \cellcolor[HTML]{FFFF00}{\textbf{90.8\%}} \\
 & AUC  & 91.8 & \cellcolor[HTML]{FFFF00}{\textbf{96.3}} & 94.7 & 95.4 \\ \hline
\multirow{4}{*}{COVID-19} & Sens & 89.1\% & \cellcolor[HTML]{FFFF00}{\textbf{96.4\%}} & 89.1\% & 89.1\% \\
 & Spec & 72.1\% & 72.1\% & 88.4\% & \cellcolor[HTML]{FFFF00}{\textbf{93.0\%}} \\
 & Prec & 80.3\% & 81.5\% & 90.7\% & \cellcolor[HTML]{FFFF00}{\textbf{94.2\%}} \\
 & F1 & 84.5\% & 88.3\% & 89.9\% & \cellcolor[HTML]{FFFF00}{\textbf{91.6\%}} \\ \hline
\end{tabular}%
\end{table}

\subsection{Classification Performance for the Public Dataset}
\label{sec:public-dataset-models}
We conducted experiments to evaluate COVID-VR for the ternary classification model (\cv vs. CAP vs. Normal) in the validation set and the three test sets from the \PUBDATASET dataset (here unified in a single test set). All models use the ResNet101 backbone network and TF6 for volume rendering. We compared COVID-VR against the winner of the competition (TheSaviours)~\cite{Chaudhary2021}, and two state-of-the-art methods, DeCoVNet~\cite{Wang2020} and COVNet~\cite{Li2020}.

\begin{table*}[!tbh]
\centering
\footnotesize
\setlength\tabcolsep{0.7pt} 
\caption{Ternary classification results by class in the Public \PUBDATASET. Performance metrics are in percentage.}
\label{tab:all-results-micro}
\resizebox{\textwidth}{!}{  
\begin{tabular}{clcc|cccccccccccc|}
\cline{5-16}
\multicolumn{1}{l}{{ \textbf{}}}	& \multicolumn{1}{c}{\textbf{}}	& \multicolumn{1}{l}{{ \textbf{}}}	& { \textbf{}}	& \multicolumn{12}{c|}{{ By Classes}} \\ \cline{3-16} 
\multicolumn{1}{l}{}	& \multicolumn{1}{c|}{}	& \multicolumn{2}{c|}{General}	& \multicolumn{4}{c|}{{ COVID-19}}	& \multicolumn{4}{c|}{{ Normal}}	& \multicolumn{4}{c|}{{ CAP}}	 \\ 
\cline{3-16} 
\multicolumn{1}{l}{}	& \multicolumn{1}{l|}{{ }}	& \multicolumn{1}{l}{Accu.}	& \multicolumn{1}{l|}{{ Kappa}}	& \multicolumn{1}{l}{{ Preci.}}	& \multicolumn{1}{l}{{ Sensi.}}	& \multicolumn{1}{l}{{ Speci}}	& \multicolumn{1}{l|}{{ F-score}}	& \multicolumn{1}{l}{{ Preci.}}	& \multicolumn{1}{l}{{ Sensi.}}	& \multicolumn{1}{l}{{ Speci.}}	& \multicolumn{1}{l|}{{ F-score}}	& \multicolumn{1}{l}{{ Preci.}}	& \multicolumn{1}{l}{{ Sensi.}}	& \multicolumn{1}{l}{{ Speci.}}	& \multicolumn{1}{l|}{{ F-score}} \\ \hline
\multicolumn{1}{|c|}{}	& \multicolumn{1}{l|}{{ TF 1}}	& { 79.6}	& { 63.7}	& { 80.3}	& { 89.1}	& { 72.1}	& \multicolumn{1}{c|}{{ 84.5}}	& { 71.4}	& { 83.3}	& {89.2}	& \multicolumn{1}{c|}{{ 76.9}}	& { \cellcolor[HTML]{FFFF00}{\textbf{100.0}}}	& { 47.4}	& { \cellcolor[HTML]{FFFF00}{\textbf{100.0}}}	& { 64.3} \\
\multicolumn{1}{|c|}{}	& \multicolumn{1}{l|}{{ TF 2}}	& { 85.7}	& {74.0}	& { 81.5}	& { \cellcolor[HTML]{FFFF00}{\textbf{96.4}}}	& {72.1}	& \multicolumn{1}{c|}{{ 88.3}}	& { \cellcolor[HTML]{FFFF00}{\textbf{95.7}}}	& {91.7}	& { \cellcolor[HTML]{FFFF00}{\textbf{98.6}}}	& 
\multicolumn{1}{c|}{{ 93.6}}	& { 90.0}	& { 47.4}	& { 98.7}	& { 62.1} \\
\multicolumn{1}{|c|}{}	& \multicolumn{1}{l|}{{ TF 3}}	& { 87.8}	& { 79.2}	& { 90.7}	& { 89.1}	& { 88.4}	& \multicolumn{1}{c|}{{ 89.9}}	& { 85.2}	& { 95.8}	& {94.6}	& \multicolumn{1}{c|}{{ 90.2}}	& {82.4}	& { 73.7}	& {96.2}	& { 77.8}	\\
\multicolumn{1}{|c|}{}	& \multicolumn{1}{l|}{{ TF 4}}	& { 88.8}	& { 80.8}	& { 90.9}	& { 90.9}	& { 88.4}	& \multicolumn{1}{c|}{{ 90.9}}	& { 85.2}	& { 95.8}	& { 94.6}	& \multicolumn{1}{c|}{{ 90.2}}	& { 87.5}	& { 73.7}	& { 97.5}	& { 80.0}	 \\
\multicolumn{1}{|c|}{}	& \multicolumn{1}{l|}{{ TF 5}}	& { 85.7}	& { 75.4}	& { 87.5}	& { 89.1}	& { 83.7}	& \multicolumn{1}{c|}{{ 88.3}}	& { 82.1}	& { 95.8}	& { 93.2}	& \multicolumn{1}{c|}{{ 88.5}}	& { 85.7}	& { 63.2}	& { 97.5}	& { 72.7}	\\
\multicolumn{1}{|c|}{\multirow{-6}{*}{\begin{tabular}[c]{@{}c@{}}Transfer\\ Function\\ Comparison\end{tabular}}}	& \multicolumn{1}{l|}{{ TF 6}}	& { \cellcolor[HTML]{FFFF00}{\textbf{90.8}}}	& { \cellcolor[HTML]{FFFF00}{\textbf{84.6}}}	& { \cellcolor[HTML]{FFFF00}{\textbf{94.2}}}	& { \textbf{89.1}}	& { \cellcolor[HTML]{FFFF00}{\textbf{93.0}}}	& \multicolumn{1}{c|}{{ \cellcolor[HTML]{FFFF00}{\textbf{91.6}}}}	& { 85.7}	& { \cellcolor[HTML]{FFFF00}{\textbf{100.0}}}	& { 94.6}	& \multicolumn{1}{c|}{{ \cellcolor[HTML]{FFFF00}{\textbf{92.3}}}}	& { 88.9}	& { \cellcolor[HTML]{FFFF00}{\textbf{84.2}}}	& {97.5}	& { \cellcolor[HTML]{FFFF00}{\textbf{86.5}}}	\\ \hline
\multicolumn{1}{|c|}{{ }}	& \multicolumn{1}{l|}{{ VGG16}}	& { 88.8}	& {81.0}	& { 90.7}	& { \cellcolor[HTML]{FFFF00}{\textbf{89.1}}}	& {88.4}	& \multicolumn{1}{c|}{{89.9}}	& { \cellcolor[HTML]{FFFF00}{\textbf{88.9}}}	& { \cellcolor[HTML]{FFFF00}{\textbf{100.0}}}	& { \cellcolor[HTML]{FFFF00}{\textbf{95.9}}}	& \multicolumn{1}{c|}{{ \cellcolor[HTML]{FFFF00}{\textbf{94.1}}}}	& { 82.4}	& { 73.7}	& { 96.2}	& { 77.8}	 \\
\multicolumn{1}{|c|}{{ }}	& \multicolumn{1}{l|}{{ DenseNet121}}	& { 87.8}	& { 79.7}	& {94.0}	& { 85.5}	& { \cellcolor[HTML]{FFFF00}{\textbf{93.0}}}	& \multicolumn{1}{c|}{{89.5}}	& { 82.8}	& { \cellcolor[HTML]{FFFF00}{\textbf{100.0}}}	& { 93.2}	& \multicolumn{1}{c|}{{ 90.6}}	& { 78.9}	& { 78.9}	& { 94.9}	& { 78.9}	 \\
\multicolumn{1}{|c|}{{ }}	& \multicolumn{1}{l|}{{ EfficientNetB2}}	& { 86.7}	& { 78.0}	& { 92.0}	& { 83.6}	& { 90.7}	& \multicolumn{1}{c|}{{87.6}}	& { 85.7}	& { \cellcolor[HTML]{FFFF00}{\textbf{100.0}}}	& { 94.6}	& \multicolumn{1}{c|}{{ 92.3}}	& { 75.0}	& { 78.9}	& { 93.7}	& { 76.9}	 \\
\multicolumn{1}{|c|}{\multirow{-4}{*}{{ \begin{tabular}[c]{@{}c@{}}Architecture\\ Comparison\end{tabular}}}}	& \multicolumn{1}{l|}{{ ResNet101}}	& { \cellcolor[HTML]{FFFF00}{\textbf{90.8}}}	& { \cellcolor[HTML]{FFFF00}{\textbf{84.6}}}	& { \cellcolor[HTML]{FFFF00}{\textbf{94.2}}}	& { \cellcolor[HTML]{FFFF00}{\textbf{89.1}}}	& { \cellcolor[HTML]{FFFF00}{\textbf{93.0}}}	& \multicolumn{1}{c|}{{ \cellcolor[HTML]{FFFF00}{\textbf{91.6}}}}	& { 85.7}	& { \cellcolor[HTML]{FFFF00}{\textbf{100.0}}}	& {94.6}	& \multicolumn{1}{c|}{{92.3}}	& { \cellcolor[HTML]{FFFF00}{\textbf{88.9}}}	& { \cellcolor[HTML]{FFFF00}{\textbf{84.2}}}	& { \cellcolor[HTML]{FFFF00}{\textbf{97.5}}}	& { \cellcolor[HTML]{FFFF00}{\textbf{86.5}}}	 \\ \hline
\multicolumn{1}{|c|}{{ }}	& \multicolumn{1}{l|}{{ DeCovNet}}	& { 67.3}	& { 44.9}	& { 74.5}	& { 74.5}	& { 67.4}	& \multicolumn{1}{c|}{{ 74.5}}	& {71.4}	& { 41.7}	& { \cellcolor[HTML]{FFFF00}{\textbf{94.6}}}	& \multicolumn{1}{c|}{{ 52.6}}	& { 51.7}	& { 78.9}	& { 82.3}	& { 62.5}	 \\
\multicolumn{1}{|c|}{{ }}	& \multicolumn{1}{l|}{{ COVNet}}	& { 77.6}	& { 63.2}	& { 85.4}	& { 74.5}	& { 83.7}	& \multicolumn{1}{c|}{{ 79.6}}	& { 66.7}	& { 83.3}	& { 86.5}	& \multicolumn{1}{c|}{{ 74.1}}	& { 75.0}	& { 78.9}	& { 93.7}	& { 76.9}	 \\
\multicolumn{1}{|c|}{{ }}	& \multicolumn{1}{l|}{{ TheSaviours}}	& { 74.5}	& { 58.8}	& { 86.7}	& { 70.9}	& { 86.0}	& \multicolumn{1}{c|}{{ 78.0}}	& { 51.4}	& { 75.0}	& {77.0}	& \multicolumn{1}{c|}{{61.0}}	& { \cellcolor[HTML]{FFFF00}{\textbf{88.9}}}	& { \cellcolor[HTML]{FFFF00}{\textbf{84.2}}}	& { \cellcolor[HTML]{FFFF00}{\textbf{97.5}}}	& { \cellcolor[HTML]{FFFF00}{\textbf{86.5}}}	 \\
\multicolumn{1}{|c|}{\multirow{-4}{*}{{ \begin{tabular}[c]{@{}c@{}}Method\\ Comparison\\ (Train /\\ Validation)\end{tabular}}}}	& \multicolumn{1}{l|}{{ COVID-VR}}	& { \cellcolor[HTML]{FFFF00}{\textbf{90.8}}}	& { \cellcolor[HTML]{FFFF00}{\textbf{84.6}}}	& { \cellcolor[HTML]{FFFF00}{\textbf{94.2}}}	& { \cellcolor[HTML]{FFFF00}{\textbf{89.1}}}	& { \cellcolor[HTML]{FFFF00}{\textbf{93.0}}}	& \multicolumn{1}{c|}{{ \cellcolor[HTML]{FFFF00}{\textbf{91.6}}}}	& { \cellcolor[HTML]{FFFF00}{\textbf{85.7}}}	& { \cellcolor[HTML]{FFFF00}{\textbf{100.0}}}	& { \cellcolor[HTML]{FFFF00}{\textbf{94.6}}}	& \multicolumn{1}{c|}{{ \cellcolor[HTML]{FFFF00}{\textbf{92.3}}}}	& { \cellcolor[HTML]{FFFF00}{\textbf{88.9}}}	& { \cellcolor[HTML]{FFFF00}{\textbf{84.2}}}	& { \cellcolor[HTML]{FFFF00}{\textbf{97.5}}}	& { \cellcolor[HTML]{FFFF00}{\textbf{86.5}}}	\\ \hline
\multicolumn{1}{|c|}{{ }}	& \multicolumn{1}{l|}{{ DeCovNet}}	& { 52.2}	& { 30.3}	& { 48.8}	& { 57.1}	& { 61.8}	& \multicolumn{1}{c|}{{ 52.6}}	& { 77.8}	& { 20.0}	& { 96.4}	& \multicolumn{1}{c|}{{ 31.8}}	& { 50.0}	& { \cellcolor[HTML]{FFFF00}{\textbf{100.0}}}	& { 71.4}	& { 66.7}	\\
\multicolumn{1}{|c|}{{ }}	& \multicolumn{1}{l|}{{ COVNet}}	& { 67.8}	& { 50.0}	& { 60.0}	& { 77.1}	& { 67.3}	& \multicolumn{1}{c|}{{ 67.5}}	& { 74.1}	& { 57.1}	& { 87.3}	& \multicolumn{1}{c|}{{ 64.5}}	& { 77.8}	& { 70.0}	& { 94.3}	& { 73.7}	 \\
\multicolumn{1}{|c|}{{ }}	& \multicolumn{1}{l|}{{ TheSaviours}}	& { \cellcolor[HTML]{FFFF00}{\textbf{90.0}}}	& { \cellcolor[HTML]{FFFF00}{\textbf{84.6}}}	& { \cellcolor[HTML]{FFFF00}{\textbf{90.9}}}	& { 85.7}	& { \cellcolor[HTML]{FFFF00}{\textbf{94.5}}}	& \multicolumn{1}{c|}{{ 88.2}}	& { 89.2}	& { \cellcolor[HTML]{FFFF00}{\textbf{94.3}}}	& { 97.1}	& \multicolumn{1}{c|}{{ \cellcolor[HTML]{FFFF00}{\textbf{90.0}}}}	& { \cellcolor[HTML]{FFFF00}{\textbf{90.0}}}	& { 90.0}	& { \cellcolor[HTML]{FFFF00}{\textbf{90.0}}}	& { \cellcolor[HTML]{FFFF00}{\textbf{94.4}}}	\\
\multicolumn{1}{|c|}{\multirow{-4}{*}{{ \begin{tabular}[c]{@{}c@{}}Method\\ Comparison\\ (Test)\end{tabular}}}}	& \multicolumn{1}{l|}{{ COVID-VR}}	& { 86.7}	& { 79.7}	& { 89.2}	& { \cellcolor[HTML]{FFFF00}{\textbf{94.3}}}	& { 92.7}	& 
\multicolumn{1}{c|}{{ \cellcolor[HTML]{FFFF00}{\textbf{91.7}}}}	& { \cellcolor[HTML]{FFFF00}{\textbf{96.4}}}	& { 77.1}	& { \cellcolor[HTML]{FFFF00}{\textbf{98.2}}}	& \multicolumn{1}{c|}{{ 85.7}}	& { 72.0}	& {90.0}	& { \cellcolor[HTML]{FFFF00}\textbf{{90.0}}}	& { 80.0}	 \\ \hline
\end{tabular}
}
\end{table*}

Table \ref{tab:all-results-micro} 
summarizes the results. 
COVID-VR achieved the highest overall performance in the validation set. COVID-VR accuracy was 90.8\% in contrast to 77.6\% obtained with COVNet, which ranked second. In terms of AUC score, COVID-VR achieved 95.4 while COVNet achieved 89.8. COVID-VR also had the best predictive power for \cv cases, reaching high and balanced sensitivity and specificity values. When comparing the CAP and Normal cases, COVID-VR had a 100\% sensitivity and 94\% specificity for the Normal class and 84.2\% sensitivity and 97.5\% specificity for the CAP class - in both cases surpassing the competing strategies.

COVID-VR achieved an accuracy of 86.7\% in the test set, while TheSaviours correctly classified 90.0\% of the test instances. 
COVID-VR was the best method for detecting \cv cases (94.3\% sensitivity), keeping high specificity (92.7\%), and F1-score (91.7\%).  In contrast, TheSaviours achieved the highest F1-score for Normal and CAP cases. The COVID-VR model is trained with labels at the patient level, while the approach presented by TheSaviours~\cite{Chaudhary2021} trains a model by exploring labels at the slice level. We reach similar results despite using a coarser-grained annotation in CT scans.

\begin{figure*}[!tb]
    \centering
        \includegraphics[width=0.49\linewidth]{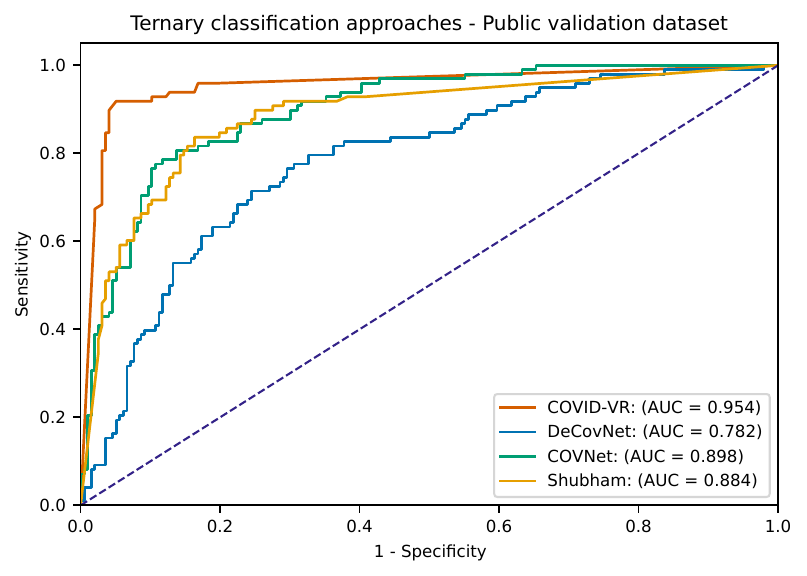}
	\includegraphics[width=0.49\linewidth]{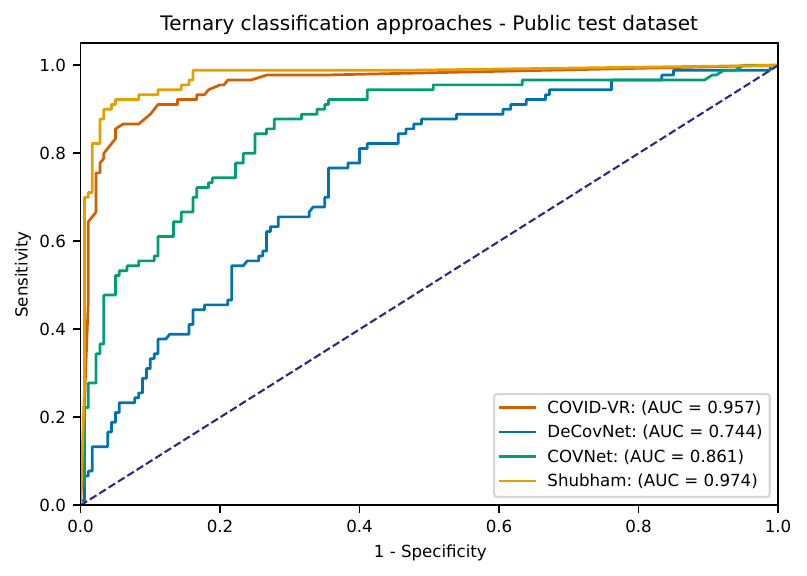}
    \caption{Micro-average ROC curves for the ternary classification task using the public dataset: validation set (left) and test set (right).} 
    \label{fig:roc-ternaryclass-publicdata}
\end{figure*}

Figure~\ref{fig:roc-ternaryclass-publicdata} compares the micro-average ROC curves for the ternary classification models using the validation test (top) and the test set (bottom). COVID-VR has the best performance in the validation set, notably improving the true positive rate (\ie sensitivity) for false-positive rates ranging from 0 to 0.3. Considering the test set, the performance of the TheSaviours model improves in experiments with the validation set and surpasses COVID-VR (\ie 97.4 vs. 95.7) in the AUC score. Nonetheless, we highlight that COVID-VR had the most stable performance between validation and test sets, despite the clinical and technical differences introduced in the CT images from the SPGC-COVID Test Set \cite{heidarian2021robust}. Finally, we note that COVID-VR had an accuracy close to that reported by the first place of the competition like IITDelhi~\cite{Pratyush2021} with 88.9\%, LLSCP~\cite{Zaifeng2021} with 87.8\%, and UniSheff\_EEE~\cite{Xue2021} with 85.56\%. The results for these approaches were not included in the table due to the lack of public code to reproduce the experiments. 



\begin{figure*}[!tb]
    \centering
     \includegraphics[width=0.49\linewidth]{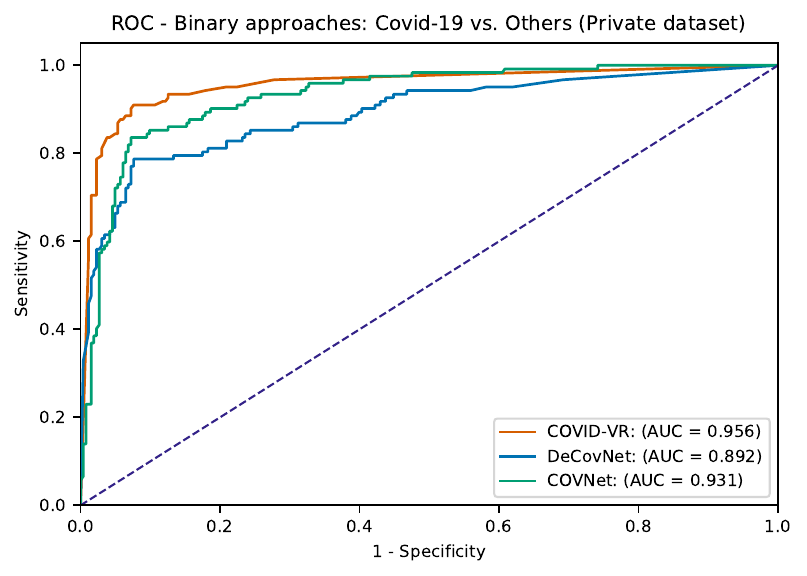}
     \includegraphics[width=0.49\linewidth]{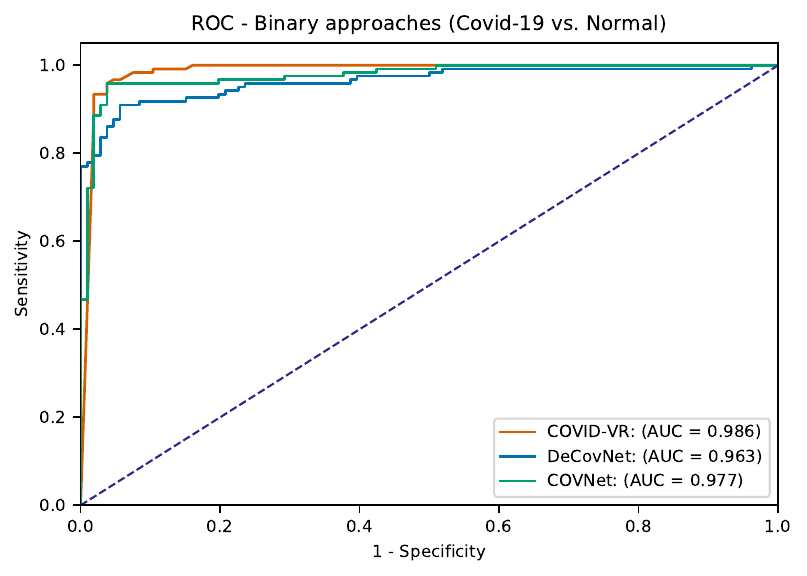}
    \caption{ROC curves for the \cv against the class others (left) and normal (right) in the private dataset.}
     \label{fig:ROC-comparison-hh-binary-approaches}
\end{figure*}




\subsection{Classification Performance for the Private Dataset}
We analyzed the performance of the methods for a binary classification task, \cv vs. non-\cv, using the private datasets. The experiments considered two definitions for the negative class: in the first, we merged the Negative, Indeterminate, and Atypical classes into a unique non-\cv class, and in the second, we only considered the original negative class (\ie Negative for pneumonia) as the classifiers non-\cv class. In both cases, the Typical classification was considered the positive class (\ie \cv). Performance assessment was based on a 5-fold CV, with the configuration of the same folds for COVID-VR, DeCoVNet, and COVNet. COVID-VR obtained the best results in all metrics, with 92.2\% of accuracy and 95.6\% of AUC for this binary classification task (Table \ref{tab:ComparisonApproachesInHH}). In comparing \cv vs. Normal classification, COVID-VR obtained accuracy and F1-score similar to COVNet, but with higher Sensitivity and AUC scores (Table \ref{tab:ComparisonApproachesInHHCovNor}). The ROC curves are given in Figure~\ref{fig:ROC-comparison-hh-binary-approaches}, that shows that in both cases COVID-VR has the best AUC scores.  


\begin{table}[!tb]
    \caption{COVID-19 vs. Others task results. Training and validation technique in the private (HMV+HCPA) dataset }
    \label{tab:ComparisonApproachesInHH}
    \centering
\begin{tabular}{cccc}
\hline
Metrics & COVID-VR & DeCovNet & COVNet \\ \hline
Acc                               & \cellcolor[HTML]{FFFF00}{\textbf{92.2\%}}   & 87.8\%                             & 89.4\%                           \\
Sens                              & \cellcolor[HTML]{FFFF00}{\textbf{83.6\%}}   & 78.7\%                             & 83.6\%                           \\
Spec                              & \cellcolor[HTML]{FFFF00}{\textbf{96.2\%}}   & 92.0\%                             & 92.0\%                           \\
F1                                & \cellcolor[HTML]{FFFF00}{\textbf{87.2\%}}   & 80.3\%                             & 83.3\%                           \\
AUC                               & \cellcolor[HTML]{FFFF00}{\textbf{95.6}}     & 89.2                               & 93.1                             \\ \hline
\end{tabular}
\end{table}

\begin{table}[!tb]
\caption{COVID-19 vs. Normal task results. Training and validation technique in the private (HMV+HCPA) dataset }
\label{tab:ComparisonApproachesInHHCovNor}
\centering
\begin{tabular}{cccc}
\hline
Metrics & COVID-VR & DeCovNet & COVNet \\ \hline
Acc                               & \cellcolor[HTML]{FFFF00}{\textbf{96.1\%}}   & 92.5\%                             & \cellcolor[HTML]{FFFF00}{\textbf{96.1\%}} \\
Sens                              & \cellcolor[HTML]{FFFF00}{\textbf{96.7\%}}   & 91.0\%                             & 95.9\%                           \\
Spec                              & 95.3\%                             & 94.3\%                             & \cellcolor[HTML]{FFFF00}{\textbf{96.2\%}} \\
F1                                & \cellcolor[HTML]{FFFF00}{\textbf{96.3\%}}   & 92.9\%                             & \cellcolor[HTML]{FFFF00}\textbf{96.3\%} \\
AUC                               & \cellcolor[HTML]{FFFF00}{\textbf{98.6}}     & 96.3                               & 97.7                             \\ \hline
\end{tabular}
\end{table}


\begin{figure*}[!tb]
    \centering
    \includegraphics[width=1.0\linewidth]{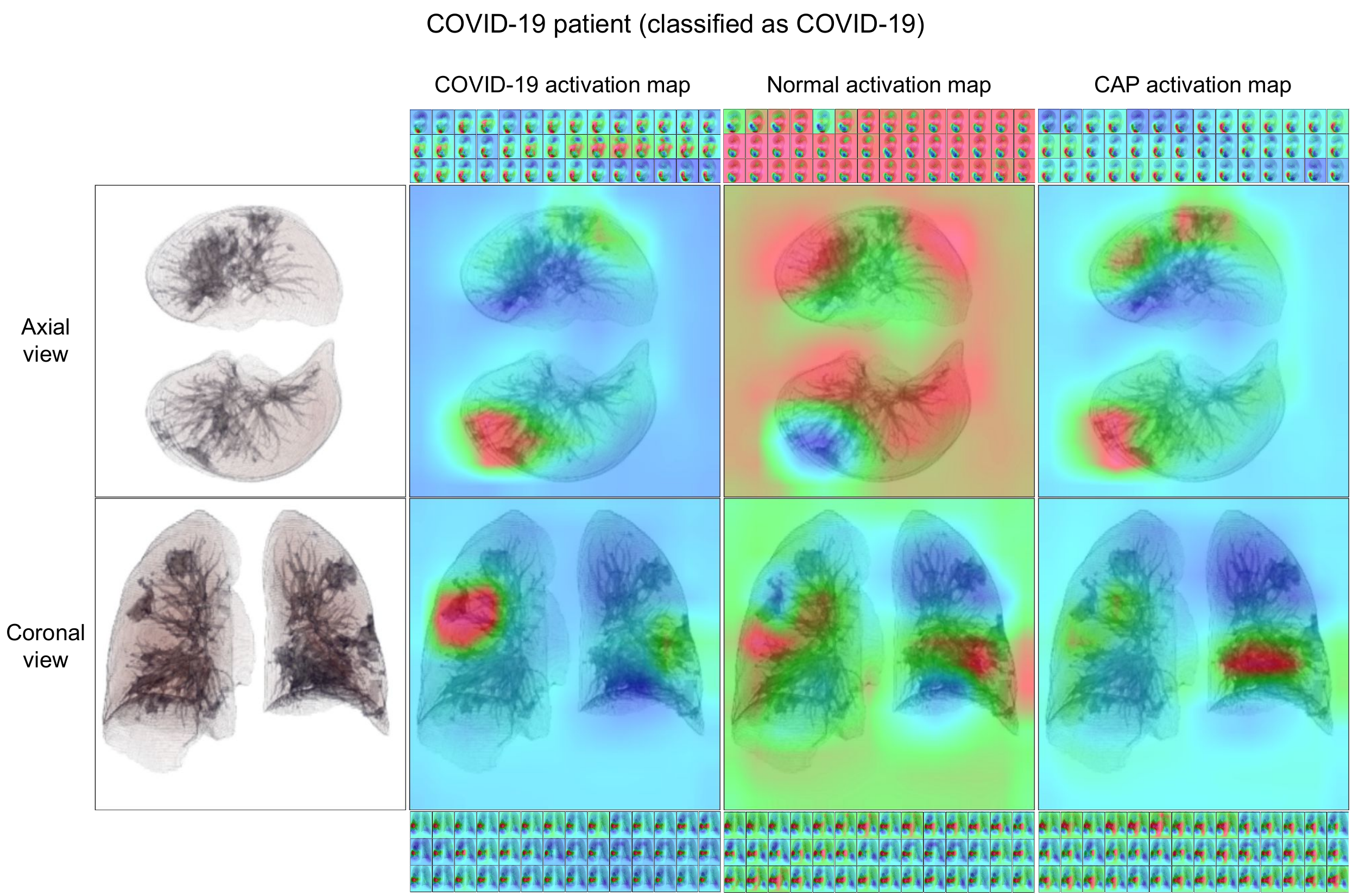}
    \caption{Grad-CAM visual activations for COVID-VR using a \cv patient from COVID-CT-MD dataset. The first column displays the volume rendering image.
    The second column shows heatmaps indicating which areas from the input image most activated the model for the class \cv, leading it to correctly classify that patient. In contrast, the second and third column show the heatmaps of activation maps for Normal and CAP class respectively. The first and last rows show thumbnails of all the activation maps.} 
%
    \label{fig:gradcam_covid}
\end{figure*}

\subsection{Visual Explanation of Activations using Grad-CAM}
We generated Grad-CAM images~\cite{Selvaraju_2017_ICCV} to explain the learning of the COVID-VR model. Figure~\ref{fig:gradcam_covid} shows the last convolutional layer activation heatmaps for the COVID-VR ternary model (\cv, Normal, and CAP) over the axial and coronal view of a \cv patient. The GRAD-CAM uses the default jet colormap. Since The COVID-VR model classifies 42 images (rotated by small angles) per patient and view, each image generates an activation map for each class, illustrated in Figure \ref{fig:gradcam_covid} as thumbnail images. The axial and coronal views show the mean of activation maps for each class from the central point of view of the camera.


The Grad-CAM heatmaps show what most influenced the classification. The \cv class corresponds to red areas with lesions in the axial and coronal views. The Normal class activates almost the entire image, avoiding one of the main lesions from \cv activation map in axial and coronal views and other visible lesions in the coronal view. Almost the entire lung is activated in the CAP class but less intense than in the Normal class.\\


\begin{table*}[!tbh]
\centering
\setlength\tabcolsep{0.7pt} 
\caption{Micro and macro average values for the Public Dataset ICASSP-SPGC Competition. Performance metrics are in percentage}
\label{tab:all-results-macro}
\begin{tabular}{cl|ccccc|ccccc|}
\cline{3-12}
\multicolumn{1}{l}{}    & \multicolumn{1}{c|}{} & \multicolumn{5}{c|}{{Micro-average}} & \multicolumn{5}{c|}{{Macro-average}} \\ \cline{3-12} 
\multicolumn{1}{l}{}    & \multicolumn{1}{l|}{} & \multicolumn{1}{l}{{Preci.}} & \multicolumn{1}{l}{{Sensi.}} & \multicolumn{1}{l}{{Speci.}} & \multicolumn{1}{l|}{{F-score}}   & \multicolumn{1}{l|}{{AUC}} & \multicolumn{1}{l}{{Preci.}} & \multicolumn{1}{l}{{Sensi.}} & \multicolumn{1}{l}{{Speci.}} & \multicolumn{1}{l|}{{F-score}}   & \multicolumn{1}{l|}{{AUC}} \\ \hline
\multicolumn{1}{|c|}{} & \multicolumn{1}{l|}{{TF 1}}   & {81.9}   & {79.6}   & {81.7}   & \multicolumn{1}{c|}{{78.7}} & {91.8}  & {86.1}   & {86.2}   & {93.1}   & \multicolumn{1}{c|}{{86.0}} & {94.1} \\
\multicolumn{1}{|c|}{} & \multicolumn{1}{l|}{{TF 2}}   & {86.6}   & {85.7}   & {83.7}   & \multicolumn{1}{c|}{{84.5}} & {\cellcolor[HTML]{FFFF00}{\textbf{96.3}}} & {89.1}   & {78.5}   & {89.8}   & \multicolumn{1}{c|}{{81.3}} & {\cellcolor[HTML]{FFFF00}{\textbf{95.2}}}\\
\multicolumn{1}{|c|}{} & \multicolumn{1}{l|}{{TF 3}}   & {87.7}   & {87.8}   & {91.4}   & \multicolumn{1}{c|}{{87.6}} & {94.7} & {83.9}   & {73.3}   & {87.1}   & \multicolumn{1}{c|}{{75.2}} & {90.2} \\
\multicolumn{1}{|c|}{} & \multicolumn{1}{l|}{{TF 4}}   & {88.8}   & {88.8}   & {91.7}   & \multicolumn{1}{c|}{{88.6}} & {95.2} & {87.9}   & {86.8}   & {93.5}   & \multicolumn{1}{c|}{{87.0}} & {94.7} \\
\multicolumn{1}{|c|}{} & \multicolumn{1}{l|}{{TF 5}}   & {85.8}   & {85.7}   & {88.7}   & \multicolumn{1}{c|}{{85.3}} & {95.0} & {85.1}   & {82.7}   & {95.1}   & \multicolumn{1}{c|}{{83.2}} & {95.2} \\
\multicolumn{1}{|c|}{\multirow{-4}{*}{{\begin{tabular}[c]{@{}c@{}}Transfer\\Function\\Comparison\end{tabular}}}} & \multicolumn{1}{l|}{{TF 6}}   & {\cellcolor[HTML]{FFFF00}{\textbf{91.1}}}   & {\cellcolor[HTML]{FFFF00}{\textbf{90.8}}}   & {\cellcolor[HTML]{FFFF00}{\textbf{94.3}}}   & \multicolumn{1}{c|}{{{\cellcolor[HTML]{FFFF00}{\textbf{90.8}}}}} & {95.4} & {\cellcolor[HTML]{FFFF00}{\textbf{89.6}}}   & {\cellcolor[HTML]{FFFF00}{\textbf{91.1}}}   & {\cellcolor[HTML]{FFFF00}{\textbf{95.0}}}   & \multicolumn{1}{c|}{{{\cellcolor[HTML]{FFFF00}{\textbf{90.1}}}}} & {\cellcolor[HTML]{FFFF00}{\textbf{95.2}}}  \\ \hline
\multicolumn{1}{|c|}{} & \multicolumn{1}{l|}{{VGG16}}   & {88.6}   & {88.8}   & {91.7}   & \multicolumn{1}{c|}{{88.6}} & {95.4} & {87.3}   & {87.6}   & {93.5}   & \multicolumn{1}{c|}{{87.3}} & {94.9} \\
\multicolumn{1}{|c|}{} & \multicolumn{1}{l|}{{DenseNet121}}   & {88.3}   & {87.8}   & {93.4}   & \multicolumn{1}{c|}{{87.7}} & {\cellcolor[HTML]{FFFF00}{\textbf{96.5}}} & {85.2}   & {88.1}   & {93.7}   & \multicolumn{1}{c|}{{86.3}} & {\cellcolor[HTML]{FFFF00}{\textbf{95.7}}} \\
\multicolumn{1}{|c|}{} & \multicolumn{1}{l|}{{EfficientNetB2}}   & {87.2}   & {86.7}   & {92.2}   & \multicolumn{1}{c|}{{86.7}} & {95.1} & {84.2}   & {87.5}   & {93.0}   & \multicolumn{1}{c|}{{85.6}} & {94.0} \\
\multicolumn{1}{|c|}{\multirow{-4}{*}{{\begin{tabular}[c]{@{}c@{}}Architecture\\ Comparison\end{tabular}}}} & \multicolumn{1}{l|}{{ResNet101}}   & {\cellcolor[HTML]{FFFF00}{\textbf{91.1}}}   & {\cellcolor[HTML]{FFFF00}{\textbf{90.8}}}   & {\cellcolor[HTML]{FFFF00}{\textbf{94.3}}}   & \multicolumn{1}{c|}{{{\cellcolor[HTML]{FFFF00}{\textbf{90.8}}}}} & {95.4} & {\cellcolor[HTML]{FFFF00}{\textbf{89.6}}}   & {\cellcolor[HTML]{FFFF00}{\textbf{91.1}}}   & {\cellcolor[HTML]{FFFF00}{\textbf{95.0}}}   & \multicolumn{1}{c|}{{{\cellcolor[HTML]{FFFF00}{\textbf{90.1}}}}} & {95.2} \\ \hline
\multicolumn{1}{|c|}{} & \multicolumn{1}{l|}{{DeCovNet}}   & {69.3}   & {67.3}   & {76.9}   & \multicolumn{1}{c|}{{66.8}} & {78.2} & {65.9}   & {65.0}   & {81.4}   & \multicolumn{1}{c|}{{63.2}} & {76.8} \\
\multicolumn{1}{|c|}{} & \multicolumn{1}{l|}{{COVNet}}   & {78.8}   & {77.5}   & {86.3}   & \multicolumn{1}{c|}{{77.7}} & {89.8} & {75.7}   & {78.9}   & {88.0}   & \multicolumn{1}{c|}{{76.9}} & {89.3} \\
\multicolumn{1}{|c|}{} & \multicolumn{1}{l|}{{TheSaviours}}   & {78.5}   & {74.5}   & {86.0}   & \multicolumn{1}{c|}{{75.5}} & {88.4} & {75.7}   & {76.7}   & {86.8}   & \multicolumn{1}{c|}{{75.2}} & {87.7} \\
\multicolumn{1}{|c|}
{\multirow{-4}{*}
{
{
\begin{tabular}[c]{@{}c@{}}Method\\ Comparison\\ (Train /\\ Validation)\end{tabular}}}} & 
\multicolumn{1}{l|}{{COVID-VR}}   & {\cellcolor[HTML]{FFFF00}{\textbf{91.1}}}   & {\cellcolor[HTML]{FFFF00}{\textbf{90.8}}}   & {\cellcolor[HTML]{FFFF00}{\textbf{94.3}}}   & 
\multicolumn{1}{c|}{{{\cellcolor[HTML]{FFFF00}{\textbf{90.8}}}}} & {\cellcolor[HTML]{FFFF00}{\textbf{95.4}}} & {\cellcolor[HTML]{FFFF00}{\textbf{89.6}}}   & {\cellcolor[HTML]{FFFF00}{\textbf{91.1}}}   & {\cellcolor[HTML]{FFFF00}{\textbf{95.0}}}   & 
\multicolumn{1}{c|}{{{\cellcolor[HTML]{FFFF00}{\textbf{90.1}}}}} & {\cellcolor[HTML]{FFFF00}{\textbf{95.2}}}\\ \hline
\multicolumn{1}{|c|}{} & \multicolumn{1}{l|}{{DeCovNet}}   & {60.3}   & {52.2}   & {77.4}   & \multicolumn{1}{c|}{{47.6}} & {78.7} & {58.9}   & {59.0}   & {76.5}   & \multicolumn{1}{c|}{{50.4}} & {78.7} \\
\multicolumn{1}{|c|}{} & \multicolumn{1}{l|}{{COVNet}}   & {69.4}   & {67.7}   & {81.1}   & \multicolumn{1}{c|}{{67.7}} & {86.1} & {70.6}   & {68.1}   & {83.0}   & \multicolumn{1}{c|}{{68.6}} & {86.6} \\
\multicolumn{1}{|c|}{} & \multicolumn{1}{l|}{{TheSaviours}}   & {\cellcolor[HTML]{FFFF00}{\textbf{90.0}}}   & {\cellcolor[HTML]{FFFF00}{\textbf{90.0}}}   & {\cellcolor[HTML]{FFFF00}{\textbf{94.4}}}   & \multicolumn{1}{c|}{{{\cellcolor[HTML]{FFFF00}{\textbf{90.0}}}}} & {\cellcolor[HTML]{FFFF00}{\textbf{97.4}}} & {\cellcolor[HTML]{FFFF00}{\textbf{90.0}}}   & {\cellcolor[HTML]{FFFF00}{\textbf{90.0}}}   & {\cellcolor[HTML]{FFFF00}{\textbf{94.8}}}   & \multicolumn{1}{c|}{{{\cellcolor[HTML]{FFFF00}{\textbf{90.0}}}}} & {\cellcolor[HTML]{FFFF00}{\textbf{98.0}}} \\
\multicolumn{1}{|c|}{\multirow{-4}{*}{{\begin{tabular}[c]{@{}c@{}}Method\\ Comparison\\ (Test)\end{tabular}}}} & \multicolumn{1}{l|}{{COVID-VR}}   & {88.2}   & {86.7}   & {94.2}  & \multicolumn{1}{c|}{{86.8}} & {95.7} & {85.9}   & {87.1}   & {93.6}  & \multicolumn{1}{c|}{{85.8}} & {96.1} \\ \hline
\end{tabular}
\end{table*}
\section{Discussion}
In summary, the main results showed that the COVID-VR architecture reached an accuracy of 90.8\% and an F1-score of 90.8\% in ternary classification using the COVID-CT-MD~\cite{Afshar2021} public dataset. The binary classification of \cv vs. others (Negative, Indeterminate, and Atypical CT images) achieved an accuracy of 92.2\% with an F1-score of 87.2\%. Finally, COVID-VR had an accuracy of 96.1\% with an F1-score of 96.3\% in the binary classification \cv vs. Normal (Negative) task using \cv class as the positive class in the private datasets given by partner hospitals.

The experiments suggest that COVID-VR achieves the goal of learning to recognize typical \cv patterns in chest CT images in comparison to other competing strategies. The COVID-VR model can help specialists in the \cv diagnosis by performing a binary classification
that identifies or discards typical cases of \cv.
Although TheSaviours model leads the performance for the ternary classification model, COVID-VR has competitive results and does not require labeling lesions in slides. In summary, COVID-VR reveals the potential of using images from the exterior of the lungs in comparison to slices used in traditional approaches.

To allow our proposal to be reproduced and compared against other proposals, we made the COVID-VR model available at {\footnotesize \ttfamily \url{https://github.com/covid-vr/covid-vr-docker}}.

\section{Conclusion and Future Work}

In this work, we introduced COVID-VR, a novel 3D Volume Rendering classification architecture  designed for classifying pulmonary diseases using volume-rendering images of the lungs. The architecture consists of three main modules: segmentation, volume rendering, and classification. The segmentation module removes non-lung material from the input chest CT scan, while the volume rendering module generates lung images from various angles. Unlike slice-based approaches that rely on images from specific CT slices (axial, coronal, or sagittal), the volume rendering technique offers a comprehensive view of the entire lung in each image, overcoming potential occlusions. Transparency is used to render the inner structures of the lung, and images are generated from angles that capture different views of the lung. Finally, the classification module employs a ResNet architecture to classify the volume rendering images into two or three classes (COVID-19, CAP and normal). 

To evaluate the effectiveness of our approach, we conducted experiments using a publicly available COVID-CT-MD dataset and a private dataset from partner hospitals. The classification results are compared against competing strategies, demonstrating that COVID-VR achieves competitive classification results without requiring the labeling of lesions in individual slides.

We recognize that there is still room for improvement in the COVID-VR model. One area to explore further is the generation of transfer functions using deep learning methods. Additionally, we aim to investigate the application of volume-rendered images in other classification scenarios, expanding the potential of our approach beyond pulmonary diseases.

\section*{Acknowledgments}
This work was partially financed by the Coordenação de Aperfeiçoamento de Pessoal de Nível Superior - Brasil (CAPES) - Finance Code 001, FAPERGS 20/2551-0000254-3 and CNPq 140313/2017-6. 

\bibliographystyle{unsrt}  
\bibliography{paper}

\begin{thebibliography}{10}

\bibitem{Hernandez-Huerta2020-yx}
Mar{\'\i}a~Teresa Hern{\'a}ndez-Huerta, Laura P{\'e}rez-Campos~Mayoral,
  Luis~Manuel S{\'a}nchez~Navarro, Gabriel Mayoral-Andrade, Eduardo
  P{\'e}rez-Campos~Mayoral, Edgar Zenteno, and Eduardo P{\'e}rez-Campos.
\newblock Should {RT-PCR} be considered a gold standard in the diagnosis of
  {COVID-19}?
\newblock {\em J Med Virol}, 93(1):137--138, July 2020.

\bibitem{Pecoraro2021-gn}
Valentina Pecoraro, Antonella Negro, Tommaso Pirotti, and Tommaso Trenti.
\newblock Estimate false-negative {RT-PCR} rates for {SARS-CoV-2}. a systematic
  review and meta-analysis.
\newblock {\em Eur J Clin Invest}, 52(2):e13706, December 2021.

\bibitem{HASSAN2022105123}
Haseeb Hassan, Zhaoyu Ren, Huishi Zhao, Shoujin Huang, Dan Li, Shaohua Xiang,
  Yan Kang, Sifan Chen, and Bingding Huang.
\newblock {Review and classification of AI-enabled COVID-19 CT imaging models
  based on computer vision tasks}.
\newblock {\em Computers in Biology and Medicine}, 141:105123, 2022.

\bibitem{Tabik2020}
S.~Tabik, A.~Gómez-Ríos, J.~L. Martín-Rodríguez, M.~Sevillano-García, I.
  Rey-Area, D.~Charte, E.~Guirado, J.~L. Suárez, et~al.
\newblock {COVIDGR Dataset and COVID-SDNet Methodology for Predicting COVID-19
  Based on Chest X-Ray Images}.
\newblock {\em IEEE Journal of Biomedical and Health Informatics},
  24(12):3595--3605, 2020.

\bibitem{Roberts2021}
Michael Roberts, Derek Driggs, Matthew Thorpe, Julian Gilbey, Michael Yeung,
  et~al.
\newblock {Common Pitfalls And Recommendations For Using Machine Learning To
  Detect And Prognosticate For COVID-19 Using Chest Radiographs and CT Scans}.
\newblock {\em Nature Machine Intelligence}, 3(3):199--217, 3 2021.

\bibitem{Kwee2020ChestKnow}
Thomas~C. Kwee and Robert~M. Kwee.
\newblock {Chest CT in COVID-19: What the radiologist needs to know}.
\newblock {\em Radiographics}, 40(7):1848--1865, 2020.

\bibitem{ALHASAN2021101933}
Mustafa Alhasan and Mohamed Hasaneen.
\newblock {Digital imaging, technologies and artificial intelligence
  applications during COVID-19 pandemic}.
\newblock {\em Computerized Medical Imaging and Graphics}, 91:101933, 2021.

\bibitem{RESNET}
Kaiming He, Xiangyu Zhang, Shaoqing Ren, and Jian Sun.
\newblock {Deep Residual Learning for Image Recognition}.
\newblock In {\em 2016 IEEE Conference on Computer Vision and Pattern
  Recognition (CVPR)}, pages 770--778, 2016.

\bibitem{Afshar2021}
Parnian Afshar, Shahin Heidarian, Nastaran Enshaei, Farnoosh Naderkhani,
  Moezedin~Javad Rafiee, Anastasia Oikonomou, Faranak~Babaki Fard, Kaveh
  Samimi, Konstantinos~N. Plataniotis, and Arash Mohammadi.
\newblock {COVID-CT-MD, COVID-19 computed tomography scan dataset applicable in
  machine learning and deep learning}.
\newblock {\em Scientific Data}, 8, 12 2021.

\bibitem{Jin2020}
Cheng Jin, Weixiang Chen, Yukun Cao, Zhanwei Xu, Zimeng Tan, Xin Zhang, Lei
  Deng, Chuansheng Zheng, Jie Zhou, Heshui Shi, and Jianjiang Feng.
\newblock {Development and evaluation of an artificial intelligence system for
  COVID-19 diagnosis}.
\newblock {\em Nature Communications}, 11(1), 2020.

\bibitem{Wang2020}
Xinggang Wang, Xianbo Deng, Qing Fu, Qiang Zhou, Jiapei Feng, Hui Ma, Wenyu
  Liu, and Chuansheng Zheng.
\newblock {A Weakly-Supervised Framework for COVID-19 Classification and Lesion
  Localization From Chest CT}.
\newblock {\em IEEE Transactions on Medical Imaging}, 39(8):2615--2625, 2020.

\bibitem{Li2020}
Lin Li, Lixin Qin, Zeguo Xu, Youbing Yin, Xin Wang, Bin Kong, Junjie Bai,
  Yi~Lu, Zhenghan Fang, Qi~Song, Kunlin Cao, Daliang Liu, Guisheng Wang,
  Qizhong Xu, Xisheng Fang, Shiqin Zhang, Juan Xia, and Jun Xia.
\newblock {Using Artificial Intelligence to Detect COVID-19 and
  Community-acquired Pneumonia Based on Pulmonary CT: Evaluation of the
  Diagnostic Accuracy}.
\newblock {\em Radiology}, 296:E65--E71, 2020.

\bibitem{He2020}
Xuehai He, Xingyi Yang, Shanghang Zhang, Jinyu Zhao, Yichen Zhang, Eric Xing,
  and Pengtao Xie.
\newblock {Sample-Efficient Deep Learning for COVID-19 Diagnosis Based on CT
  Scans}.
\newblock {\em medRxiv}, 2020.

\bibitem{Wang2020b}
Linda Wang, Zhong~Qiu Lin, and Alexander Wong.
\newblock {COVID-Net: a tailored deep convolutional neural network design for
  detection of COVID-19 cases from chest X-ray images}.
\newblock {\em Scientific Reports}, 10, 12 2020.

\bibitem{Gozes2020}
Ophir Gozes, Ma'ayan Frid-Adar, Hayit Greenspan, Patrick~D. Browning, Huangqi
  Zhang, Wenbin Ji, Adam Bernheim, and Eliot Siegel.
\newblock {Rapid AI development cycle for the coronavirus (COVID-19) pandemic:
  Initial results for automated detection {\&} patient monitoring using deep
  learning CT image analysis}, 2020.

\bibitem{AMYAR2020104037}
Amine Amyar, Romain Modzelewski, Hua Li, and Su~Ruan.
\newblock {Multi-task deep learning based CT imaging analysis for COVID-19
  pneumonia: Classification and segmentation}.
\newblock {\em Computers in Biology and Medicine}, 126:104037, 2020.

\bibitem{IBRAHIM2021104348}
Dina~M. Ibrahim, Nada~M. Elshennawy, and Amany~M. Sarhan.
\newblock {Deep-chest: Multi-classification deep learning model for diagnosing
  COVID-19, pneumonia, and lung cancer chest diseases}.
\newblock {\em Computers in Biology and Medicine}, 132:104348, 2021.

\bibitem{SERTE2021104306}
Sertan Serte and Hasan Demirel.
\newblock {Deep learning for diagnosis of COVID-19 using 3D CT scans}.
\newblock {\em Computers in Biology and Medicine}, 132:104306, 2021.

\bibitem{DASILVEIRA2023103775}
Thiago~L.T. {da Silveira}, Paulo~G.L. Pinto, Thiago~S. Lermen, and Cláudio~R.
  Jung.
\newblock Omnidirectional 2.5d representation for covid-19 diagnosis using
  chest cts.
\newblock {\em Journal of Visual Communication and Image Representation},
  91:103775, 2023.

\bibitem{Chaudhary2021}
Shubham Chaudhary, Sadbhawna Sadbhawna, Vinit Jakhetiya, Badri~N Subudhi,
  Ujjwal Baid, and Sharath~Chandra Guntuku.
\newblock {Detecting COVID-19 and Community Acquired Pneumonia Using Chest CT
  Scan Images With Deep Learning}.
\newblock In {\em ICASSP 2021 - IEEE International Conference on Acoustics,
  Speech and Signal Processing}, pages 8583--8587, 2021.

\bibitem{Pratyush2021}
Pratyush Garg, Rishabh Ranjan, Kamini Upadhyay, Monika Agrawal, and Desh
  Deepak.
\newblock {Multi-Scale Residual Network for COVID-19 Diagnosis Using CT-Scans}.
\newblock In {\em ICASSP 2021 - IEEE International Conference on Acoustics,
  Speech and Signal Processing (ICASSP)}, pages 8558--8562, 2021.

\bibitem{Zaifeng2021}
Zaifeng Yang, Yubo Hou, Zhenghua Chen, Le~Zhang, and Jie Chen.
\newblock {A Multi-Stage Progressive Learning Strategy for Covid-19 Diagnosis
  Using Chest Computed Tomography with Imbalanced Data}.
\newblock In {\em ICASSP 2021 - IEEE International Conference on Acoustics,
  Speech and Signal Processing (ICASSP)}, pages 8578--8582, 2021.

\bibitem{Xue2021}
Shuohan Xue and Charith Abhayaratne.
\newblock {Covid-19 Diagnostic Using 3d Deep Transfer Learning for
  Classification of Volumetric Computerised Tomography Chest Scans}.
\newblock In {\em ICASSP 2021 - IEEE International Conference on Acoustics,
  Speech and Signal Processing (ICASSP)}, pages 8573--8577, 2021.

\bibitem{Bougourzi2021}
Fares Bougourzi, Riccardo Contino, Cosimo Distante, and Abdelmalik Taleb-Ahmed.
\newblock {CNR-IEMN: A Deep Learning Based Approach to Recognise Covid-19 from
  CT-Scan}.
\newblock In {\em ICASSP 2021 - IEEE International Conference on Acoustics,
  Speech and Signal Processing (ICASSP)}, pages 8568--8572, 2021.

\bibitem{Li2021}
Bingyang Li, Qi~Zhang, Yinan Song, Zhicheng Zhao, Zhu Meng, and Fei Su.
\newblock {Diagnosing COVID-19 from CT Images Based on an Ensemble Learning
  Framework}.
\newblock In {\em ICASSP 2021 - IEEE International Conference on Acoustics,
  Speech and Signal Processing (ICASSP)}, pages 8563--8567, 2021.

\bibitem{arens2010survey}
Stephan Arens and Gitta Domik.
\newblock A survey of transfer functions suitable for volume rendering.
\newblock In {\em VG@ Eurographics}, pages 77--83, 2010.

\bibitem{jonsson2014survey}
Daniel Jönsson, Erik Sundén, Anders Ynnerman, and Timo Ropinski.
\newblock A survey of volumetric illumination techniques for interactive volume
  rendering.
\newblock {\em Computer Graphics Forum}, 33(1):27--51, 2014.

\bibitem{kniss2002multidimensional}
Joe Kniss, Gordon Kindlmann, and Charles Hansen.
\newblock Multidimensional transfer functions for interactive volume rendering.
\newblock {\em IEEE Transactions on visualization and computer graphics},
  8(3):270--285, 2002.

\bibitem{zhang2011volume}
Qi~Zhang, Roy Eagleson, and Terry~M Peters.
\newblock Volume visualization: a technical overview with a focus on medical
  applications.
\newblock {\em Journal of digital imaging}, 24(4):640--664, 2011.

\bibitem{Tang2020SevereCT}
Lei Tang, Xiaoyong Zhang, Yvquan Wang, and Xianchun Zeng.
\newblock {Severe COVID-19 Pneumonia: Assessing Inflammation Burden with
  Volume-rendered Chest CT}.
\newblock {\em Radiology: Cardiothoracic Imaging}, 2(2):e200044, 2020.

\bibitem{li2020multiscale}
Guang Li, Sharon~E Fox, Brian Summa, Bihe Hu, Carola Wenk, Aibek Akmatbekov,
  Jack~L Harbert, Richard~S Vander~Heide, and J~Quincy Brown.
\newblock {Multiscale 3-dimensional pathology findings of COVID-19 diseased
  lung using high-resolution cleared tissue microscopy}.
\newblock {\em Biorxiv}, 2020.

\bibitem{Jadhav2021}
Shreeraj Jadhav, Gaofeng Deng, Marlene Zawin, and Arie~E. Kaufman.
\newblock {COVID-view: Diagnosis of COVID-19 using Chest CT}.
\newblock {\em IEEE Transactions on Visualization and Computer Graphics},
  28(1):227--237, 2022.

\bibitem{AMARA202211}
Kahina Amara, Ali Aouf, Hoceine Kennouche, A.~Oualid Djekoune, Nadia Zenati,
  Oussama Kerdjidj, and Farid Ferguene.
\newblock {COVIR: A virtual rendering of a novel NN architecture O-Net for
  COVID-19 Ct-scan automatic lung lesions segmentation}.
\newblock {\em Computers \& Graphics}, 104:11--23, 2022.

\bibitem{heidarian2021robust}
Shahin Heidarian, Parnian Afshar, Nastaran Enshaei, Farnoosh Naderkhani,
  Moezedin~Javad Rafiee, Anastasia Oikonomou, Akbar Shafiee, Faranak~Babaki
  Fard, Konstantinos~N Plataniotis, and Arash Mohammadi.
\newblock Robust automated framework for covid-19 disease identification from a
  multicenter dataset of chest ct scans.
\newblock {\em arXiv preprint arXiv:2109.09241}, 2021.

\bibitem{simpson2020radiological}
Scott Simpson, Fernando~U Kay, Suhny Abbara, Sanjeev Bhalla, Jonathan~H Chung,
  Michael Chung, Travis~S Henry, Jeffrey~P Kanne, Seth Kligerman, Jane~P Ko,
  et~al.
\newblock Radiological {S}ociety of {N}orth {A}merica expert consensus document
  on reporting chest {CT} findings related to {COVID-19}: endorsed by the
  {S}ociety of {T}horacic {R}adiology, the {A}merican {C}ollege of {R}adiology,
  and {RSNA}.
\newblock {\em Radiology: Cardiothoracic Imaging}, 2(2):e200152, 2020.

\bibitem{Sousa2019}
Azael~M. Sousa, Samuel~B. Martins, Alexandre~X. Falcão, Fabiano Reis, Ericson
  Bagatin, and Klaus Irion.
\newblock {ALTIS: A fast and automatic lung and trachea CT-image segmentation
  method}.
\newblock {\em Medical Physics}, 46:4970--4982, 2019.

\bibitem{lensink2020segmentation}
Keegan Lensink, Issam Laradji, Marco Law, Paolo~Emilio Barbano, Savvas
  Nicolaou, William Parker, and Eldad Haber.
\newblock Segmentation of pulmonary opacification in chest ct scans of covid-19
  patients.
\newblock {\em arXiv preprint arXiv:2007.03643}, 2020.

\bibitem{hofmanninger2020lungmask}
Johannes Hofmanninger, Forian Prayer, Jeanny Pan, Sebastian R{\"o}hrich, Helmut
  Prosch, and Georg Langs.
\newblock Automatic lung segmentation in routine imaging is primarily a data
  diversity problem, not a methodology problem.
\newblock {\em European Radiology Experimental}, 4(1):1--13, 2020.

\bibitem{PHNN}
Adam~P. Harrison, Ziyue Xu, Kevin George, Le~Lu, Ronald~M. Summers, and
  Daniel~J. Mollura.
\newblock Progressive and multi-path holistically nested neural networks for
  pathological lung segmentation from ct images.
\newblock In {\em Medical Image Computing and Computer Assisted Intervention
  (MICCAI)}, pages 621--629, Cham, 2017. Springer International Publishing.

\bibitem{tang2020severity}
Zhenyu Tang, Wei Zhao, Xingzhi Xie, Zheng Zhong, Feng Shi, Jun Liu, and
  Dinggang Shen.
\newblock Severity assessment of coronavirus disease 2019 (covid-19) using
  quantitative features from chest ct images.
\newblock {\em arXiv preprint arXiv:2003.11988}, 2020.

\bibitem{lu2021quantitative}
Weiping Lu, Jianguo Wei, Tingting Xu, Miao Ding, Xiaoyan Li, Mengxue He, Kai
  Chen, Xiaodan Yang, Huiyuan She, and Bingcang Huang.
\newblock {Quantitative CT for detecting COVID-19 pneumonia in suspected
  cases}.
\newblock {\em BMC infectious diseases}, 21(1):1--8, 2021.

\bibitem{wolf2004medical}
Ivo Wolf, Marcus Vetter, Ingmar Wegner, Marco Nolden, Thomas Bottger, Mark
  Hastenteufel, Max Schobinger, Tobias Kunert, and Hans-Peter Meinzer.
\newblock The medical imaging interaction toolkit (mitk): a toolkit
  facilitating the creation of interactive software by extending vtk and itk.
\newblock In {\em Medical Imaging 2004: Visualization, Image-Guided Procedures,
  and Display}, volume 5367, pages 16--27. SPIE, 2004.

\bibitem{ye2020chest}
Zheng Ye, Yun Zhang, Yi~Wang, Zixiang Huang, and Bin Song.
\newblock Chest {CT} manifestations of new coronavirus disease 2019
  ({COVID-19}): a pictorial review.
\newblock {\em European Radiology}, 30(8):4381--4389, 2020.

\bibitem{VGG}
Karen Simonyan and Andrew Zisserman.
\newblock Very deep convolutional networks for large-scale image recognition,
  2015.

\bibitem{DENSENET}
G.~Huang, Z.~Liu, L.~Van~Der Maaten, and K.~Q. Weinberger.
\newblock {Densely Connected Convolutional Networks}.
\newblock In {\em 2017 IEEE Conference on Computer Vision and Pattern
  Recognition (CVPR)}, pages 2261--2269, Los Alamitos, CA, USA, jul 2017. IEEE
  Computer Society.

\bibitem{EFFICIENTNET}
Mingxing Tan and Quoc~V. Le.
\newblock Efficientnet: Rethinking model scaling for convolutional neural
  networks, 2020.

\bibitem{Selvaraju_2017_ICCV}
Ramprasaath~R. Selvaraju, Michael Cogswell, Abhishek Das, Ramakrishna Vedantam,
  Devi Parikh, and Dhruv Batra.
\newblock Grad-cam: Visual explanations from deep networks via gradient-based
  localization.
\newblock In {\em 2017 IEEE International Conference on Computer Vision
  (ICCV)}, pages 618--626, 2017.

\end{thebibliography}

\end{document}